\documentclass[longauth]{aa}  
\usepackage{hyperref}
\usepackage[dvipsnames]{xcolor}
\usepackage{graphicx}
\usepackage{txfonts}
\usepackage{xspace}
\usepackage{booktabs}
\usepackage{makecell}
\usepackage{multirow}
\usepackage{threeparttable}
\usepackage{pifont}
\usepackage{enumitem}
\usepackage{upgreek}
\usepackage{multirow}

\newcommand{\kms}{km\,s$^{-1}$\xspace}

\newcommand{\SFRhi}{$\mathrm{SFR}_{\rm hi}$\xspace}
\newcommand{\SFRcorr}{$\mathrm{SFR}_{\rm corr}$\xspace}
\newcommand{\sSFRhi}{$\mathrm{sSFR}_{\rm hi}$\xspace}
\newcommand{\sSFRcorr}{$\mathrm{sSFR}_{\rm corr}$\xspace}

\usepackage{natbib}
\defcitealias{PaperI}{Paper~I}
\bibpunct{(}{)}{;}{a}{}{,} % to follow the A&A style
\begin{document} 

   \title{Linking neutral gas inflows and outflows to offsets in the star-forming main sequence and mass--metallicity relation}

\author{
  S.~Weng\inst{1} \and
  M.~Pieri\inst{1} \and
  A.~Saintonge\inst{2,3} \and
  D.~Scholte\inst{4} \and
  D.~Mu{\~n}oz Santos\inst{1} \and
  T.~Hu\inst{1} \and
  J.~Aguilar\inst{5} \and
  S.~Ahlen\inst{6} \and
  F.~Beutler\inst{4} \and
  D.~Bianchi\inst{7,8} \and
  D.~Brooks\inst{2} \and
  A.~Carnero Rosell\inst{9,10} \and
  F.~J.~Castander\inst{11,12} \and
  T.~Claybaugh\inst{5} \and
  A.~de la Macorra\inst{13} \and
  B.~Dey\inst{14,15} \and
  P.~Doel\inst{2} \and
  V.~A.~Fawcett\inst{16} \and
  J.~E.~Forero-Romero\inst{17,18} \and
  E.~Gazta{\~n}aga\inst{11,19,12} \and
  S.~{Gontcho A Gontcho}\inst{20} \and
  G.~Gutierrez\inst{21} \and
  A.~Kremin\inst{5} \and
  M.~Landriau\inst{5} \and
  L.~Le~Guillou\inst{22} \and
  A.~Meisner\inst{23} \and
  R.~Miquel\inst{24,25} \and
  J.~Moustakas\inst{26} \and
  S.~Nadathur\inst{19} \and
  W.~J.~Percival\inst{27,28,29} \and
  F.~Prada\inst{30} \and
  I.~P\'erez-R\`afols\inst{31} \and
  C.~Ravoux\inst{32} \and
  G.~Rossi\inst{33} \and
  R.~Ruggeri\inst{34} \and
  E.~Sanchez\inst{35} \and
  C.~Saulder\inst{36} \and
  D.~Schlegel\inst{5} \and
  J.~Silber\inst{5} \and
  M.~Siudek\inst{12,10} \and
  G.~Tarl\'{e}\inst{37} \and
  B.~A.~Weaver\inst{23} \and
  H.~Zou\inst{38}
}

\institute{
  Aix Marseille Univ, CNRS, CNES, LAM, Marseille, France
  \and
  Department of Physics \& Astronomy, University College London,
  Gower Street, London, WC1E 6BT, UK
  \and
  Max-Planck-Institut f\"ur Radioastronomie,
  Auf dem H\"ugel 69, 53121 Bonn, Germany
  \and
  Institute for Astronomy, University of Edinburgh,
  Royal Observatory, Blackford Hill, Edinburgh EH9 3HJ, UK
  \and
  Lawrence Berkeley National Laboratory,
  1 Cyclotron Road, Berkeley, CA 94720, USA
  \and
  Department of Physics, Boston University,
  590 Commonwealth Avenue, Boston, MA 02215, USA
  \and
  Dipartimento di Fisica ``Aldo Pontremoli'',
  Universit\`a degli Studi di Milano,
  Via Celoria 16, I-20133 Milano, Italy
  \and
  INAF-Osservatorio Astronomico di Brera,
  Via Brera 28, 20122 Milano, Italy
  \and
  Departamento de Astrof\'{\i}sica, Universidad de La Laguna (ULL),
  E-38206 La Laguna, Tenerife, Spain
  \and
  Instituto de Astrof\'{\i}sica de Canarias,
  C/ V\'{\i}a L\'{a}ctea, s/n,
  E-38205 La Laguna, Tenerife, Spain
  \and
  Institut d'Estudis Espacials de Catalunya (IEEC),
  c/ Esteve Terradas 1, Edifici RDIT, Campus PMT-UPC,
  08860 Castelldefels, Spain
  \and
  Institute of Space Sciences, ICE-CSIC, Campus UAB,
  Carrer de Can Magrans s/n,
  08913 Bellaterra, Barcelona, Spain
  \and
  Instituto de F\'{\i}sica,
  Universidad Nacional Aut\'{o}noma de M\'{e}xico,
  Circuito de la Investigaci\'{o}n Cient\'{\i}fica,
  Ciudad Universitaria, Cd. de M\'{e}xico,
  C.~P.~04510, M\'{e}xico
  \and
  Department of Astronomy \& Astrophysics,
  University of Toronto, Toronto, ON M5S 3H4, Canada
  \and
  Department of Physics \& Astronomy and Pittsburgh Particle Physics,
  Astrophysics, and Cosmology Center (PITT PACC),
  University of Pittsburgh, 3941 O'Hara Street,
  Pittsburgh, PA 15260, USA
  \and
  European Southern Observatory,
  Karl-Schwarzschild-Stra{\ss}e 2,
  85748 Garching bei M\"unchen, Germany
  \and
  Departamento de F\'isica, Universidad de los Andes,
  Cra. 1 No. 18A-10, Edificio Ip,
  CP 111711, Bogot\'a, Colombia
  \and
  Observatorio Astron\'omico, Universidad de los Andes,
  Cra. 1 No. 18A-10, Edificio H,
  CP 111711, Bogot\'a, Colombia
  \and
  Institute of Cosmology and Gravitation,
  University of Portsmouth, Dennis Sciama Building,
  Portsmouth, PO1 3FX, UK
  \and
  University of Virginia, Department of Astronomy,
  Charlottesville, VA 22904, USA
  \and
  Fermi National Accelerator Laboratory,
  PO Box 500, Batavia, IL 60510, USA
  \and
  Sorbonne Universit\'{e}, CNRS/IN2P3,
  Laboratoire de Physique Nucl\'{e}aire et de Hautes Energies (LPNHE),
  FR-75005 Paris, France
  \and
  NSF NOIRLab, 950 N. Cherry Ave., Tucson, AZ 85719, USA
  \and
  Instituci\'{o} Catalana de Recerca i Estudis Avan\c{c}ats,
  Passeig de Llu\'{\i}s Companys 23,
  08010 Barcelona, Spain
  \and
  Institut de F\'{\i}sica d'Altes Energies (IFAE),
  The Barcelona Institute of Science and Technology,
  Edifici Cn, Campus UAB,
  08193 Bellaterra (Barcelona), Spain
  \and
  Department of Physics and Astronomy, Siena University,
  515 Loudon Road, Loudonville, NY 12211, USA
  \and
  Department of Physics and Astronomy, University of Waterloo,
  200 University Ave W, Waterloo, ON N2L 3G1, Canada
  \and
  Perimeter Institute for Theoretical Physics,
  31 Caroline St. North, Waterloo, ON N2L 2Y5, Canada
  \and
  Waterloo Centre for Astrophysics, University of Waterloo,
  200 University Ave W, Waterloo, ON N2L 3G1, Canada
  \and
  Instituto de Astrof\'{i}sica de Andaluc\'{i}a (CSIC),
  Glorieta de la Astronom\'{i}a, s/n,
  E-18008 Granada, Spain
  \and
  Departament de F\'isica, EEBE,
  Universitat Polit\`ecnica de Catalunya,
  c/Eduard Maristany 10,
  08930 Barcelona, Spain
  \and
  Universit\'{e} Clermont-Auvergne, CNRS, LPCA,
  63000 Clermont-Ferrand, France
  \and
  Department of Physics and Astronomy, Sejong University,
  209 Neungdong-ro, Gwangjin-gu,
  Seoul 05006, Republic of Korea
  \and
  Queensland University of Technology,
  School of Chemistry \& Physics,
  George St, Brisbane 4001, Australia
  \and
  CIEMAT, Avenida Complutense 40,
  E-28040 Madrid, Spain
  \and
  Max Planck Institute for Extraterrestrial Physics,
  Gie{\ss}enbachstra{\ss}e 1,
  85748 Garching, Germany
  \and
  University of Michigan,
  500 S. State Street, Ann Arbor, MI 48109, USA
  \and
  National Astronomical Observatories,
  Chinese Academy of Sciences,
  A20 Datun Road, Chaoyang District,
  Beijing 100101, P.~R.~China
}
   \date{}

% \abstract{}{}{}{}{} 
% 5 {} token are mandatory
 
  \abstract
  % context heading (optional)
  % {} leave it empty if necessary  
    {
    Gas inflows and outflows regulate galaxy growth, but direct observational links between measured gas flows and galaxy scaling relations remain limited.
    Using approximately 6,000 star-forming galaxies with down-the-barrel \ion{Na}{i}~D absorption from DESI DR2, we examine how systems with detected neutral-gas inflows and outflows populate the star-forming main sequence (SFMS) and mass--metallicity relation (MZR).    
    Inflow and outflow hosts are compared to stellar-mass- and redshift-matched controls, and to SFMS and MZR fits derived from galaxies without detected gas flows.
    Outflow hosts ($v_{\rm flow} \leq -50$ \kms) show enhanced specific star formation rates (sSFRs) by $\sim$0.25--0.40 dex and elevated central metallicities by $\sim$0.04--0.06 dex in the lower-redshift sample.
    Slow inflow hosts ($0 < v_{\rm flow} < 100$ \kms) have similarly enhanced sSFR, by $\sim$0.20--0.30 dex, but no significant metallicity offset, while fast inflow hosts ($v_{\rm flow} \geq 100$ \kms) show weaker SFR enhancement and modestly lower metallicities.
    Together, these trends support a regulator picture in which neutral gas flows trace different phases of the baryon cycle.
    Slow inflow hosts lie above the SFMS relation, potentially consistent with accretion sustaining enhanced star formation without producing strong central metallicity dilution.
    This may indicate that inflowing gas is already metal-enriched or has mixed or enriched over extended timescales.
    By contrast, outflow hosts lie near the upper 1$\sigma$ SFMS envelope, consistent with feedback acting to regulate subsequent growth.
    Gas-flow hosts also show small but systematic offsets in the narrow 4000 \AA\ break strength ($D_n4000$) relative to controls matched in redshift, stellar mass and SFR.
    Our results provide observational evidence that neutral gas flows are associated with population-level offsets from the SFMS and MZR, consistent with a baryon-cycle contribution to scaling-relation scatter.
    }
    \titlerunning{How gas flows drive scatter in scaling relations}
    \authorrunning{Weng et al.}
   \maketitle

%-------------------------------------------------------------------
%% References to add
% https://iopscience.iop.org/article/10.1088/0004-637X/746/1/108 Metallicity changes due to mergers
% galaxy size, metallicity https://arxiv.org/abs/2606.13124
%-------------------------------------------------------------------

\section{Introduction}
% General
The tight correlations between stellar mass ($M_*$), star-formation rate (SFR) and gas-phase metallicity ($Z$) are among the fundamental scaling relations governing galaxy evolution. 
The star-forming main sequence \citep[SFMS;][]{Brinchmann2004, Whitaker2012} relates stellar mass to star formation rate, while the mass--metallicity relation (MZR) connects stellar mass to gas-phase metallicity \citep{Tremonti2004, Kewley2008}. 
Their suggested interdependence is captured by the Fundamental Metallicity Relation (FMR), in which metallicity depends on both stellar mass and SFR \citep{Ellison2008MZR, Mannucci2010}. 
Together, these relations reflect the coupled evolution of gas supply, star formation and chemical enrichment.

The slope and normalisation of these relations describe the average growth of the galaxy population, whereas their scatter contains information about shorter-term variations in the baryon cycle.
Inflows, outflows and recycled gas can move galaxies away from the mean relations by changing their gas supply, star-formation activity and chemical enrichment \citep{PerouxHowk2020, DiCesare2026}.

Gas-regulator models interpret these offsets as temporary departures from an approximate equilibrium between accretion, star formation and feedback.
In this framework, enhanced accretion first increases the gas reservoir and can dilute the metals in the ISM, then fuels elevated star formation, before gas consumption and feedback-driven outflows act to restore equilibrium \citep[e.g.][]{Bouche2010, Lilly2013, Tacchella2016}.
Cosmological simulations indicate that burstiness, halo assembly history, feedback duty cycles and fluctuations in inflow rate can all contribute to the scatter \citep{Forbes2014, Matthee2017, delucia2020, vanLoon2021}.
Their individual contributions remain difficult to isolate because these processes are coupled, operate on overlapping timescales and can produce similar changes in global galaxy properties.

% Observations of scatter, add references/content
In observations, the physical origin of the scatter is usually inferred indirectly through correlations with gas fraction, morphology, environment and other stellar properties. 
Galaxies above the SFMS tend to be more gas rich \citep{Saintonge2016}, while metallicity residuals at fixed stellar mass appear anti-correlated with SFR \citep{Ellison2008MZR, Mannucci2010} and gas content \citep{Bothwell2013, Lara-Lopez2013, Hughes2013, Brown2018}. 
Both trends are expected when recent accretion fuels star formation and dilutes the ISM. 
Observational links between measured gas flows and offsets from these coupled scaling relations remain rare \citep{Langan2023}, largely because inflow detections have been difficult to obtain. 
Establishing this connection is essential for testing whether the gas flows invoked in equilibrium models are directly associated with the offsets observed in real galaxies.

% Down-the-barrel spectroscopy intro
Down-the-barrel spectroscopy allows us to measure gas flows directly by detecting absorbing gas against the stellar continuum of the host galaxy \citep{Phillips1993}. 
The \ion{Na}{i}~D doublet is particularly useful at low redshift because it is accessible at optical wavelengths and traces the cool neutral ISM. 
Previous studies have used \ion{Na}{i}~D and complementary tracers such as \ion{Mg}{ii} and UV absorption lines to measure outflows and, less commonly, inflows in galaxies \citep[e.g.][]{Heckman2000, Martin2005, Rupke2005aSample, Rupke2005bAnalysis, ChenSDSS2010, Rubin2010, Chisholm2015, Roberts-Borsani2019, Concas2019, Sun2024}. 
However, the lack of large samples of individual gas-flow detections has limited our ability to connect these flows to galaxy scaling relations.

% Signpost, reference to paper I
In \citet{PaperI} (hereafter \citetalias{PaperI}), we identify more than 50,000 down-the-barrel absorbers in the Dark Energy Spectroscopic Instrument (DESI) survey, revealing a substantial population of inflows alongside outflows. 
Here we use these detections to ask where galaxies hosting inflows and outflows lie relative to the SFMS and MZR, and whether metallicity differences persist after controlling for SFR.  
In \autoref{sec:DESI}, we introduce the DESI data and \autoref{sec:methods} describes the measured galaxy properties. 
The offsets of galaxies hosting gas flows from the SFMS and MZR are presented in \autoref{sec:results} and interpreted in \autoref{sec:discuss} before the main conclusions in \autoref{sec:conclusion}. 
Throughout this paper we adopt the cosmological parameters from \citet{Planck2020} and the \citet{Chabrier2003} initial mass function (IMF).

\section{The Dark Energy Spectroscopic Instrument}
\label{sec:DESI}
The Dark Energy Spectroscopic Instrument is a robotic, fibre-fed, highly multiplexed spectrograph mounted on the 4-metre Mayall Telescope at Kitt Peak National Observatory in Arizona \citep{DESI2022.KP1.Instr}. 
Its focal plane is equipped with 5000 robotic fibre positioners \citep{FocalPlane.Silber.2023}, enabling simultaneous spectroscopy of nearly 5000 targets \citep{Corrector.Miller.2023, FiberSystem.Poppett.2024}. 
DESI is carrying out an eight-year survey over $\sim$17,000~deg$^2$, ultimately targeting spectra for approximately 63 million galaxies and quasars \citep{DESI2016b.Instr}. 
The scale of the project requires a dedicated suite of survey operations and data reduction pipelines \citep{Spectro.Pipeline.Guy.2023, SurveyOps.Schlafly.2023}.

Observations began during Survey Validation in 2020, with full survey operations commencing in 2021. 
The first public data release, DESI DR1 \citep{DESI2024.I.DR1}, contains approximately 14.5 million extragalactic spectra and 4 million stellar spectra from the first year of data. 
This dataset enabled early cosmological constraints from full-shape analyses \citep{DESI2024.VII.KP7B}. 
The more recent DESI DR2 expands this to over 33 million extragalactic and 12 million stellar spectra from the first three years of observations \citep{DESI.DR2.BAO.lya, DESI.DR2.BAO.cosmo}, establishing DESI as the largest spectroscopic surveys to date. 
Cosmological analyses are continuing with the forthcoming DR3 \citep{DESI.DR2.DR2}.

In this work, we use galaxy spectra from the Bright Galaxy Survey (BGS) and Luminous Red Galaxy (LRG) target classes in DESI DR2. 
The BGS targets galaxies at $z < 0.6$ during bright-time observations and consists of BGS Bright, with $r \lesssim 19.5$ mag, and BGS Faint, which extends to $19.5 < r < 20.175$ with an additional colour--magnitude selection to maintain high redshift success \citep{BGS.TS.Hahn.2023}. 
We also include LRGs, which are selected using optical and infrared colour--magnitude cuts to identify massive luminous galaxies primarily at $0.4 < z \lesssim 1.0$ \citep{LRG.TS.Zhou.2023}. 
{Restricting both target classes to $0.002 < z < 0.6$, with the upper limit set by the DESI wavelength coverage and the lower limit imposed to minimise stellar contamination, gives approximately 12.2 million BGS galaxies and 3.4 million LRGs in DR2.}
The DESI wavelength coverage and moderate spectral resolution ($3800 < R < 5500$) therefore provide a uniquely large dataset for identifying neutral interstellar gas through down-the-barrel \ion{Na}{i}~D absorption.

\section{Gas flow and galaxy properties}
\label{sec:methods}

\subsection{Down-the-barrel absorption catalogue}
In \citetalias{PaperI} we describe in detail the construction of our down-the-barrel \ion{Na}{i}~D catalogue and here provide only a brief summary. 
We restrict the parent DESI sample to spectra with median signal-to-noise ratio (SNR) per pixel $>$5 in a rest-frame $\pm50$ \AA\ window around \ion{Na}{i}~D, yielding $\sim$6 million galaxies. 
After correcting for Galactic extinction, we normalise each spectrum using the local continuum estimated from \texttt{FastSpecFit}\footnote{\url{https://fastspecfit.readthedocs.io/en/latest/index.html}} models \citep[][Moustakas et al. in prep.]{MoustakasFastSpecFitsoftware}.
The normalised spectra are fitted with a nested hierarchy of absorption models comprising a null continuum, a systemic component and up to two additional kinematic flow components described by a partial-covering model \citep{Rupke2005aSample}. 
All models are convolved with the wavelength-dependent DESI resolution matrix and evaluated within a Bayesian framework using nested sampling, enabling model comparison via evidence ratios (also referred to as the Bayes factor). 
A two-stage Bayesian selection is applied, using an initial permissive evidence threshold to identify candidates followed by a stricter evidence cut to define the final sample. 

Several minor updates were made to the sample-selection procedure relative to \citetalias{PaperI}. 
We visually inspected every candidate detection and removed those attributable to identifiable observational or data-processing artefacts; such cases constituted approximately 2.5\% of the original sample. 
We also revisited candidates rejected by the quality cuts adopted in \citetalias{PaperI} and reinstated those for which visual inspection supported an astrophysical origin. 
Following these revisions, the final sample contains 52,696 galaxies.

\subsection{Measuring galaxy properties}
\subsubsection{Stellar mass}
Stellar masses are obtained from four sources.
We use measurements from \texttt{FastSpecFit}, which jointly fits \textit{grzW1--W4} photometry from the Legacy Survey DR9 \citep{Dey2019} and WISE \citep{Wright2010} with DESI spectra via stellar population synthesis, applying an aperture correction to account for the fibre aperture.
\texttt{FastPhot} \citep[][Moustakas et al. in prep.]{MoustakasFastSpecFitsoftware} uses the same photometry but no spectra.
Two further estimates come from the Code Investigating GALaxy Emission \citep[CIGALE;][]{BurgarellaCIGALE2005, Boquien2019}, following \citet{ZouCIGALE2024}: \texttt{CIGALE\_15} incorporates 10 pseudo-filters constructed from the DESI spectrum alongside \textit{grzW1W2} photometry, whereas \texttt{CIGALE\_5} uses the photometry alone.
The four estimates allow us to test the consistency of results across different SED fitting codes without assuming any single one is correct.
More details on each code can be found in \autoref{app:SED}. 

\subsubsection{Emission line fluxes}
% Comparison of my Bayesian method with FSF in Appendix to show it doesn't matter too much, maybe add scatter? 
We use the emission-line fluxes derived by \texttt{FastSpecFit}, which performs a joint fit to the stellar continuum and nebular emission lines. 
We require a minimum SNR of 3 for an emission line to be considered detected and included in subsequent analyses. 
We further restrict the sample to $z < 0.45$ so that all emission lines required for measuring metallicities remain within the DESI wavelength coverage. 

Emission-line fluxes are measured within the DESI fibre which has a radius of 0.75\,arcsec.
At low redshift, the fibre probes only a small physical scale, with its radius increasing from roughly 0.5 kpc at $z = 0.03$ to $\sim$3 kpc at $z = 0.25$. 
Quantities derived from these fluxes, such as SFR and metallicity, therefore reflect conditions in the central region of the galaxy and may not be representative of global values, particularly for nearby systems.
We impose a general redshift floor of $z > 0.03$ for all galaxies, ensuring the fibre subtends at least $r \sim 0.5$\,kpc so that measurements probe a physically meaningful central region.

\subsubsection{BPT diagnostics}
For reliable star-formation rate and metallicity estimates, we restrict the sample to star-forming galaxies using the Baldwin--Phillips--Terlevich (BPT) diagnostic diagrams \citep{Baldwin1981, Veilleux1987}.
We require galaxies to be classified as `SF' in at least one of the three BPT diagnostics, while excluding galaxies classified as composite or AGN in any diagram \citep{Kewley2001, Kauffmann2003, Kewley2006}.
Galaxies may therefore be classified as star-forming in one diagram but remain unclassified in others due to insufficient SNR.
Such galaxies are retained in the sample, since requiring high-SNR detections in all of [\ion{N}{ii}], [\ion{S}{ii}] and [\ion{O}{i}] would preferentially select nearby strong emission-line systems and exclude otherwise normal star-forming galaxies with weak low-ionisation emission.

\subsubsection{Dust extinction and star-formation rate}
The dust extinction is estimated from the Balmer decrement, using the observed H$\alpha$/H$\beta$ ratio and assuming an intrinsic Case~B recombination value of 2.86 for typical \ion{H}{ii} region conditions \citep{OsterbrockFerland2006}. 
We adopt the attenuation law of \citet{Cardelli1989} to correct the emission-line fluxes for dust attenuation. 
Star-formation rates are then computed from the dust-corrected H$\alpha$ luminosity using the calibration of \citet{Kennicutt1998}, converted from a \citet{Salpeter1955} IMF to a \citet{Chabrier2003} IMF using a scaling factor of 0.63.

The fibre-integrated H$\alpha$ flux does not capture emission from the entire galaxy at low redshift and hence underestimates the total SFR.
We account for this limitation with two samples.
For galaxies at $0.03 < z < 0.25$, \SFRcorr scales the fibre H$\alpha$ flux to an estimate of the total flux using the \texttt{APERCORR} variable from \texttt{FastSpecFit}.
The \SFRhi sample instead contains galaxies at $z \geq 0.25$, where the DESI fibre radius subtends at least $\sim$3\,kpc, and no aperture correction is applied.
Using both \SFRcorr and \SFRhi serves as a self-consistency check on the robustness of our results to aperture effects, especially for low-redshift galaxies where the fibre predominantly samples the nuclear region.

\subsubsection{Gas-phase metallicity}
We derive gas-phase metallicities for each star-forming galaxy using the method described in \citet{Scholte2024}.
This approach combines strong-line metallicity diagnostics to infer a single metallicity estimate. 
Gas-phase metallicities inferred from strong emission lines are subject to substantial systematic uncertainties, so our conclusions rely on relative metallicity differences measured consistently within the same calibration.
The \citet{Scholte2024} metallicity calibration was derived using DESI Bright Galaxy Survey validation and Year~1 data \citep{BGS.TS.Hahn.2023}.
It combines the $R_{23}$ calibration of \citet{Nakajima2022} with the N2 calibration of \citet{Denicolo2002} and is validated against electron-temperature ($T_e$) metallicities derived from auroral-line measurements.
We infer the metallicity posterior using Markov chain Monte Carlo sampling with \texttt{emcee} \citep{ForemanMackey2013}.
We use eight walkers and 200 steps, with a burn-in of 50 steps.
This replaces the least-squares approach adopted by \citet{Scholte2024} and allows measurement uncertainties and the intrinsic scatter of each diagnostic to be propagated into the metallicity estimates.

\subsubsection{4000 \AA\, break}
The 4000\,\AA\ break arises from the combined opacity of numerous metal
absorption lines and the crowding of high-order Balmer absorption lines at
wavelengths around and below 4000\,\AA. 
We use the narrow-band break strength, $D_n4000$ \citep{Balogh1999}, measured by \texttt{FastSpecFit} from the DESI spectrum to characterise the stellar population within the DESI fibre. 
Larger values generally correspond to older or more metal-rich stellar populations, while smaller values indicate younger or less metal-rich populations. 
As $D_n4000$ depends on both stellar age and metallicity, we use it as a relative indicator of differences in the recent stellar populations of the gas-flow and control samples rather than as a unique age measurement. 
At low redshift this diagnostic primarily describes the central region sampled by the fibre and does not constrain the stellar population at larger galactocentric radii.

\subsection{Gas flow and control sample selection}
\begin{table*}
\centering
\caption{Number of star-forming galaxies in each flow category and redshift interval. Star-forming galaxies are selected using BPT classifications and are required to have dust-corrected SFR measurements.}
\label{tab:counts}
\begin{tabular}{llcc}
\hline
\hline
\textbf{Sample} &
\textbf{Definition} &
${0.03 < z < 0.25}$ &
${0.25 \leq z < 0.45}$ \\
\hline

Control
& No detected flow
& 571\,688
& 107\,992 \\

Outflow
& $v_{\rm flow} \leq -50$~\kms
& 3\,065
& 1\,377 \\

Low-velocity blueshifted
& $-50 < v_{\rm flow} < 0$~\kms
& 281
& 20 \\

Slow inflow
& $0 < v_{\rm flow} < 100$~\kms
& 764
& 60 \\

Fast inflow
& $v_{\rm flow} \geq 100$~\kms
& 277
& 130 \\

\hline
\end{tabular}
\end{table*}

\subsubsection{Separating gas-flow samples}
To define gas flows, we first establish a systemic reference velocity.
There are three possible values: the \texttt{RedRock} pipeline redshift \citep[]{Spectro.Pipeline.Guy.2023}, the \ion{Na}{i}\,D redshift ($z_{\rm Na}$) from \citetalias{PaperI} and the emission-line redshift from \texttt{FastSpecFit} \citep{MoustakasFastSpecFitsoftware}.
We adopt $z_{\rm Na}$ because the systemic and gas-flow components are fitted simultaneously using the same \ion{Na}{i}\,D absorption profile, providing the most self-consistent reference for the gas-flow velocities.

Outflows are defined as cases where the gas flow velocity is $v_{\rm flow} \leq -50$\,\kms with respect to the systemic velocity.
For inflows, \citetalias{PaperI} found strong evidence for a population of low-velocity inflows at $v_{\rm flow} \approx 20$\,\kms, meaning a threshold analogous to the outflow criterion would exclude much of this population.
We therefore define two inflow categories: slow inflows consistent with radial transport within the disc or galactic-fountain flows ($0 < v_{\rm flow} < 100$ \kms) and fast inflows associated with gas accretion from satellites beyond the disc ($v_{\rm flow} \geq 100$ \kms). 
The remaining galaxies, with $-50 < v_{\rm flow} < 0$~\kms, are designated low-velocity blueshifted; this population likely comprises a mixture of low-velocity outflows and inflows and is therefore difficult to interpret physically. 
\autoref{tab:counts} summarises the gas-flow definitions and samples used. 

\subsubsection{Matching gas-flow and control samples}
Rather than matching individual gas-flow galaxies to their nearest neighbours, we construct control samples that reproduce the joint distributions of relevant physical and observational properties. 
We apply multidimensional binning and rejection sampling to several combinations of matching parameters and repeat the procedure over 1000 Monte Carlo realisations to account for measurement
uncertainties. 
This enables a consistent comparison between galaxies with detected gas flows and otherwise similar galaxies without detected flows.

Redshift and stellar mass form the baseline matching parameters. 
Matching in redshift accounts for redshift-dependent selection effects and variations in the physical scale subtended by the fixed angular size of the DESI fibre.

Two additional parameters may affect the detection of gas flows. Outflows are preferentially detected in more face-on galaxies \citep{ChenSDSS2010}, whereas inflows are more commonly detected in edge-on systems \citep{Rubin2012, PaperI}, motivating the inclusion of axial ratio ($b/a$). 
Detection sensitivity also depends on the SNR, which we measure within a $\pm50$\,\AA region around the \ion{Na}{i}\,D line.
As the parent sample generally has a lower SNR than the sample with detected gas flows \citepalias[see][]{PaperI}, failing to match in SNR increases the likelihood that the control sample contains galaxies hosting weak or otherwise undetected flows.

We therefore consider several matching configurations. 
For the SFMS and MZR, we use two-, three- and four-parameter matching. 
For metallicity comparisons at fixed SFR, we add SFR to the redshift and stellar-mass matching, mirroring the secondary dependence captured by the FMR. 
As the number of matched parameters increases, the number of bins used in each dimension is reduced to retain sufficient galaxies per multidimensional bin. 
The adopted parameters and numbers of bins per dimension are summarised in \autoref{tab:params}.

For each configuration, we divide the parameter space into an $n_{\rm bin}^{d}$ grid, where $n_{\rm bin}$ is the number of bins per dimension and $d$ is the number of matched parameters. 
We construct normalised multidimensional histograms for both samples and assign each control galaxy an acceptance probability proportional to the ratio of the gas-flow to control-sample density in its bin. 
This procedure is equivalent to rejection sampling: control galaxies in bins that are over-represented relative to the gas-flow sample are down-weighted, producing a matched control sample with a similar distribution to the gas-flow sample. 
We restrict all samples to $\log(M_*/\rm M_\odot) > 9.5$, as the scarcity of gas-flow galaxies at lower stellar masses prevents reliable matching and can produce poorly populated bins with unstable weights.

To propagate measurement uncertainties, we repeat the matching procedure for 1000 Monte Carlo realisations. 
Quantities with substantial measurement uncertainties, such as $M_*$, are drawn from Gaussian distributions centred on their measured values, with standard deviations set by the corresponding uncertainties. 
The same perturbation procedure is applied independently to both the gas-flow and control samples before each matching realisation. 

\begin{table}[h!]
\centering
\caption{Parameters controlled for when matching the gas-flow and control samples and the number of bins used in each dimension.}
\label{tab:params}
\begin{tabular}{llc}
\hline\hline
\textbf{Comparison} & \textbf{Matched parameters} & {$\mathbf n_{\rm bin}^d$} \\
\hline
     & $z$, $M_\star$ & $15^2$ \\
SFMS & $z$, $M_\star$, one of $\{b/a,\, \mathrm{SNR}\}$ & $10^3$ \\
     & $z$, $M_\star$, $b/a,\, \mathrm{SNR}$ & $6^4$ \\
\hline
     & $z$, $M_\star$ & $15^2$ \\
MZR  & $z$, $M_\star$, one of $\{b/a,\, \mathrm{SNR}\}$ & $10^3$ \\
     & $z$, $M_\star$, $b/a,\, \mathrm{SNR}$ & $6^4$ \\
\hline
\multirow{2}{*}{MZR $+$ SFR} 
     & $z$, $M_\star$, SFR & $10^3$ \\
     & $z$, $M_\star$, SFR, one of $\{b/a,\, \mathrm{SNR}\}$ & $6^4$ \\
\hline
\end{tabular}
\end{table}

\subsubsection{Investigating the scatter in scaling relations}
In addition to comparisons between gas-flow and matched-control samples, we measure offsets from fitted galaxy scaling relations. 
These approaches are complementary: the matched samples provide an empirical comparison at fixed observed properties, while the fitted relations place the gas-flow galaxies on a common physical scale relative to the parent population. 
We emphasise that these fits are not intended to provide precise measurements of the absolute SFMS or MZR. 
Deriving those relations robustly would require a more detailed treatment of selection effects, aperture corrections and metallicity calibration systematics than is needed here. 
Instead, we use them as internally consistent reference relations, applying the same procedure to the parent sample and to the gas-flow galaxies.

For the scaling relations, we restrict the analysis to $0.03 < z < 0.25$, since the sample becomes severely incomplete at lower stellar masses for $z > 0.25$. 
We adopt the stellar-mass completeness limits from \citet{Hahn2023}, which were derived using DESI data during the One-percent Survey, and add an extra redshift bin at $0.21 < z < 0.25$ with a stellar-mass limit of $\log M_*/\rm M_\odot > 11.0$. 
As in the matching analysis, we additionally impose a global stellar-mass cut of $\log(M_*/{\rm M_\odot})>9.5$.
Together, these cuts ensure that the fitted relations are measured over a stellar-mass range reasonably well sampled by the parent population.
We note that these completeness limits were derived for the magnitude-limited BGS Bright sample, whereas our parent sample also includes BGS Faint and LRG targets.

For the SFMS, we adopt the redshift-dependent functional form of \citet{Speagle2014}, fitting its parameters to the completeness-corrected DESI control sample. 
For the MZR, we use the functional form of \citet{Curti2020}. 
The offsets are then defined as the differences between the observed values and those predicted by the corresponding best-fitting relation.

\section{Results}
\label{sec:results}

\subsection{Star-forming main sequence}
\begin{figure*}
\centering
\includegraphics[width=\linewidth]{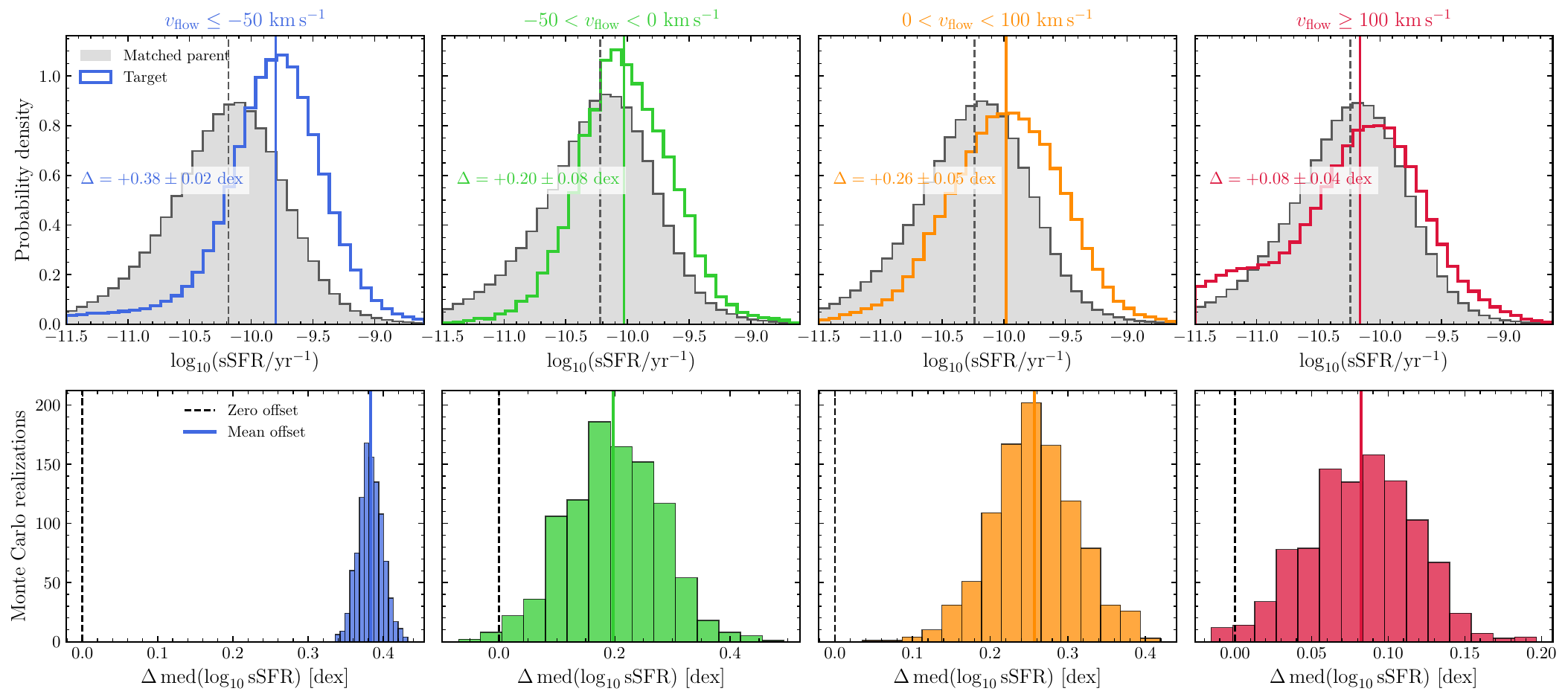}
\caption{
    Example sSFR differences between galaxies with detected gas flows and matched control samples. 
    This example uses galaxies at $z \geq 0.25$ with \SFRhi and \texttt{CIGALE\_5} stellar masses. 
    The control sample is additionally matched in axial ratio, using 10 bins along each of the three matching dimensions.
    From left to right, we show outflows ($v_{\rm flow} \leq -50$ \kms), low-velocity blueshifted gas ($-50 < v_{\rm flow} < 0$ \kms), slow inflows ($0 < v_{\rm flow} < 100$ \kms) and fast inflows ($v_{\rm flow} \geq 100$ \kms). 
    The top row shows the average sSFR probability densities after 1000 MC realisations, while the bottom row shows the distribution of median differences. 
    }
\label{fig:hist_show}
\end{figure*}

We compare the star-formation activity of gas-flow galaxies with matched controls and with a fitted SFMS.
For the \SFRhi\ sample, \autoref{fig:hist_show} shows the resulting sSFR distributions after matching in redshift, \texttt{CIGALE\_5} stellar mass and axial ratio.
The distributions of the control samples (filled histograms) and gas-flow samples (coloured histograms) are averaged over 1000 Monte Carlo realisations, with vertical lines marking their medians.
The bottom row shows the distribution of the difference in median sSFR between the gas-flow and control samples across these realisations.
We adopt the mean of this distribution as the final offset and its $1\sigma$ standard deviation as the uncertainty.
The effectiveness of the matching is demonstrated in \autoref{fig:match}, which compares the parameter distributions before and after matching for the same \SFRhi\ configuration.

In this example, outflow hosts have a median sSFR offset of $0.38 \pm 0.02$ dex relative to the control sample. 
Galaxies with low-velocity blueshifted absorbers, slow inflows and fast inflows also show elevated sSFRs, with median offsets of $0.20 \pm 0.08$, $0.26 \pm 0.05$ and $0.08 \pm 0.04$ dex, respectively. 
Among the inflow classes, the enhancement is strongest for slow inflows, while the fast-inflow offset is smaller but remains significant at the $2\sigma$ level. 
To assess the robustness of these trends, we repeat the analysis using different stellar-mass estimates, SFR measurements and combinations of matched parameters, as outlined in \autoref{app:SED}.

\begin{figure*}
\centering
\includegraphics[width=\linewidth]{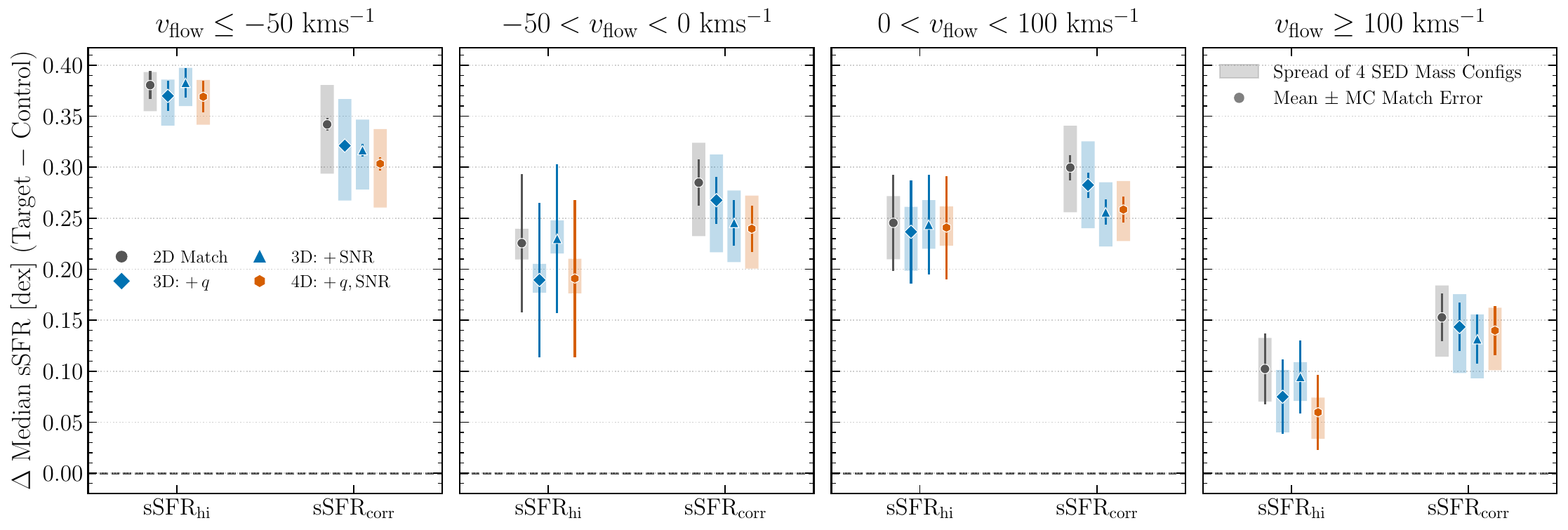}
\caption{
    Differences in sSFR between gas-flow and matched control samples for different matching variables and stellar-mass estimates. 
    The vertical bars indicate the range of results obtained using the four stellar-mass estimates. 
    Points show the mean across these estimates and error bars show the typical uncertainty of an individual measurement. 
    The vertical bars therefore reflect the systematic uncertainty associated with the choice of stellar-mass estimate, while the error bars reflect the statistical uncertainty from the finite sample size and matching process. 
    We consider two SFR measurements: \SFRhi for galaxies at $z \geq 0.25$ and aperture-corrected SFRs (\SFRcorr) for galaxies at $0.03 < z < 0.25$. 
    Colours correspond to the dimensionality of the matching, while symbols indicate the matched parameters as shown in the legend. 
        }
\label{fig:sf_diff}
\end{figure*}

\autoref{fig:sf_diff} summarises the sSFR offsets. 
The colours indicate the dimensionality of the matching procedure, corresponding to two-parameter (grey), three-parameter (blue) and four-parameter (orange) matching. 
Different symbols indicate the specific parameters included in the matching. 
For each configuration, we calculate a sample-level median sSFR offset separately using each of the four SED-based stellar-mass estimates. 
The vertical shaded region spans the minimum and maximum of these four offsets and therefore represents the sensitivity to the adopted stellar-mass estimator. 
The point marks the mean of the four offsets, while the error bar shows the typical $1\sigma$ uncertainty obtained from the Monte Carlo realisations for an individual stellar-mass estimate. 

The measurements are grouped according to the adopted sSFR indicator: \sSFRhi for galaxies at $z \geq 0.25$ and aperture-corrected SFRs (\sSFRcorr) for galaxies at $0.03 < z < 0.25$. 
\sSFRhi and \sSFRcorr agree across the gas-flow populations and matching configurations. 
The trends are therefore not strongly tied to the choice of global SFR estimator.

\begin{figure*}
\centering
\includegraphics[width=\linewidth]{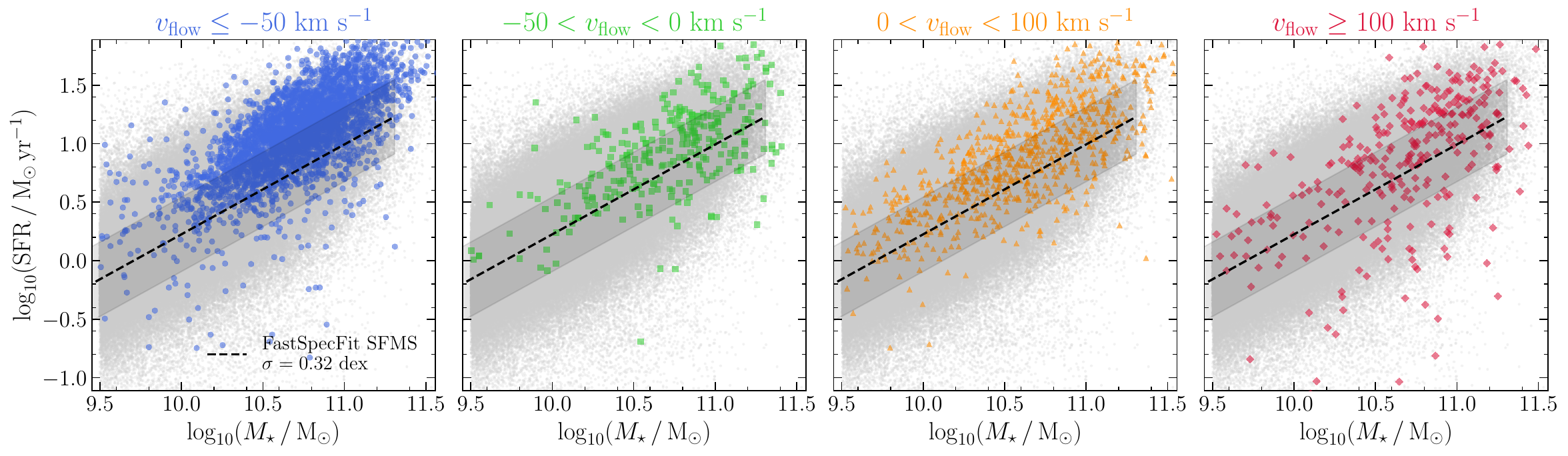}
\caption{
    The star-forming main sequence for the mean redshift of the control sample, $z \approx 0.14$, applying the mass-completeness limits from \citet{Hahn2023}. 
    Stellar masses are taken from \texttt{FastSpecFit} and the star-formation rates are aperture corrected. 
    The dashed line shows the best-fitting relation using the parameterisation of \citet{Speagle2014}, with the grey shaded region showing the intrinsic $1\sigma$ scatter. 
    Galaxies with different gas-flow velocities are shown as coloured points in each panel while grey points show the parent sample. 
}
\label{fig:sfms_fsf}
\end{figure*}

We also measure offsets from the SFMS fitted to the DESI control sample. 
An example fitted SFMS is shown in \autoref{fig:sfms_fsf} using \texttt{FastSpecFit} stellar masses and aperture-corrected SFRs. 
Galaxies hosting outflows, low-velocity blueshifted absorbers and slow inflows lie visibly above the fitted main sequence, which has a scatter of $\sigma_{\rm MS} = 0.32$ dex. 

\begin{figure}
\centering
\includegraphics[width=\linewidth]{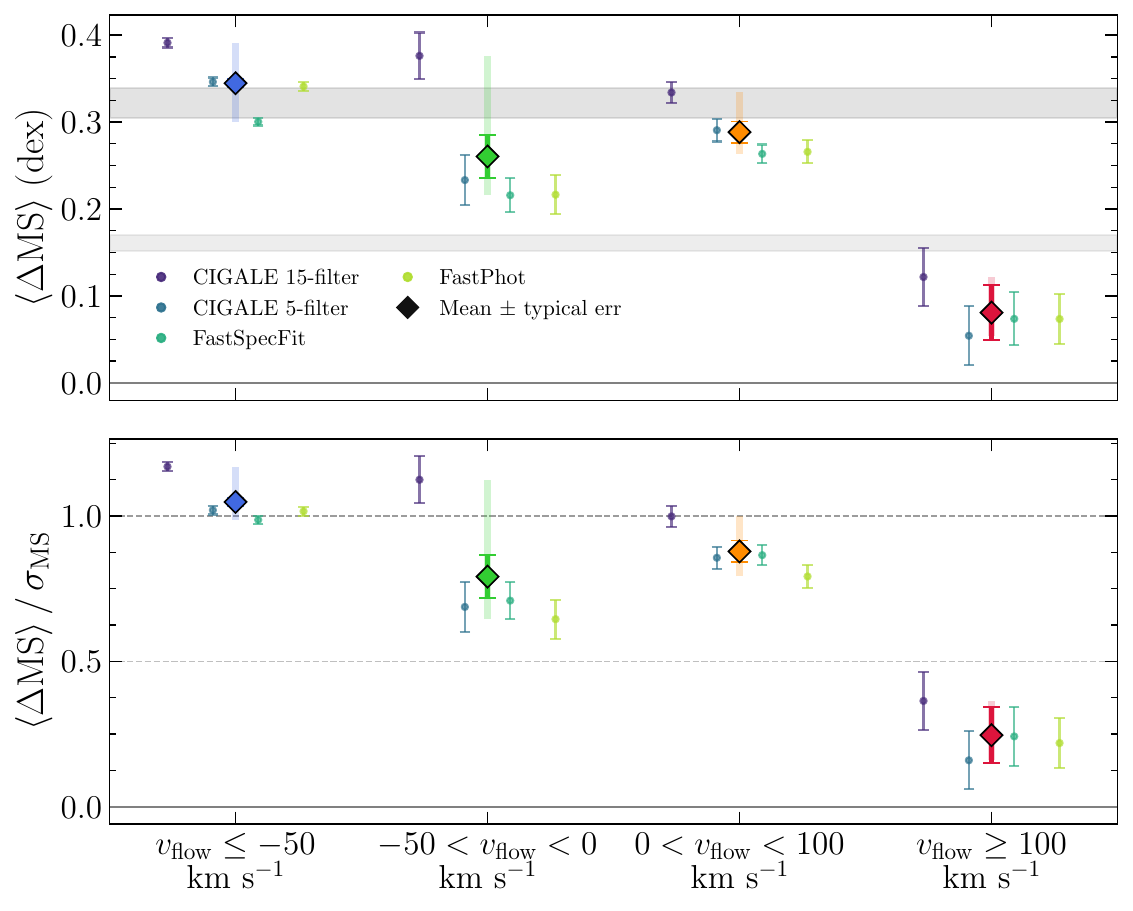}
\caption{
    Mean offset from the star-forming main sequence for galaxies hosting gas flows at different velocities. 
    Coloured circles show the results obtained using four stellar-mass estimates, while the translucent vertical bars span their full range without accounting for individual uncertainties. 
    Large diamonds mark the mean across the four estimates, with error bars indicating the typical uncertainty of an individual estimate. 
    The upper panel presents the offset in dex; the grey bands show the typical ranges corresponding to $0.5\sigma_{\rm MS}$ and $1.0\sigma_{\rm MS}$. 
    The lower panel shows the offset normalised by the measured main-sequence scatter, with dashed horizontal lines at the same values. 
    Solid grey lines indicate zero offset.
    }
\label{fig:delta_ms}
\end{figure}

We apply this method using four different stellar-mass estimates, with the resulting offsets for each gas-flow velocity bin shown in \autoref{fig:delta_ms}. 
The top panel presents the offsets in dex, while the bottom panel shows the offsets normalised by the measured SFMS scatter. 
The fitted-SFMS offsets agree with the matched-control comparison in \autoref{fig:sf_diff}.

\subsubsection{Outflows: $v_{\rm flow} \leq -50$ \kms}

Across all combinations of stellar masses, SFR measurements and matched parameters, galaxies hosting outflows consistently exhibit higher sSFRs than their matched controls. 
The \sSFRhi and \sSFRcorr enhancements agree within the systematic range associated with the different stellar-mass catalogues and span 0.25--0.40 dex (\autoref{fig:sf_diff}). 
As seen in \autoref{fig:delta_ms}, outflow hosts also lie $\sim$0.3--$0.4$ dex above the fitted SFMS, corresponding to $\sim$1$\sigma_{\rm MS}$. 
Both analyses recover the same enhancement.

As BPT-selected AGN have been removed from the star-forming sample, the observed outflows are likely driven primarily by star formation. 
The sSFR enhancement therefore reflects the host-galaxy conditions associated with detectable outflows rather than a direct consequence of the outflowing gas. 
The down-the-barrel absorption selection is more complete for stronger outflows with higher velocities \citepalias{PaperI}, which may preferentially occur in galaxies with elevated star-formation activity \citep{ChenSDSS2010}.
As the offsets remain largely unchanged after matching on axial ratio, orientation cannot explain the enhancement.

\subsubsection{Low-velocity blueshifted gas: $-50 < v_{\rm flow} < 0$ \kms}

Galaxies with low-velocity blueshifted \ion{Na}{i}\,D absorption also show enhanced sSFRs relative to their matched controls, although the offsets are smaller than for the outflow population. 
Across the matching configurations, \sSFRhi and \sSFRcorr are elevated by approximately 0.15--0.30 dex in \autoref{fig:sf_diff}. 
The smaller sample of low-velocity blueshifted absorbers at $z \geq 0.25$ leads to larger statistical uncertainties for \sSFRhi, but the broad agreement with \sSFRcorr supports the robustness of the trend. 
The fitted-SFMS offsets agree with the matched-control measurements, with this population lying $\sim$0.25 dex above the relation, or $\sim$0.7$\sigma_{\rm MS}$ (\autoref{fig:delta_ms}).

The physical origin of this class remains uncertain. 
These systems may represent the low projected-velocity tail of the outflow population, for example when an outflow is viewed closer to edge-on.
During visual inspection, we also identified some cases where inaccuracies in the \texttt{RedRock} systemic redshifts may cause genuinely redshifted absorption features to be classified as weak blueshifts. 
The intermediate sSFR offsets also favour a mixture of physical origins.

\subsubsection{Slow inflows: $0 < v_{\rm flow} < 100$ \kms}
Galaxies hosting slow inflows show enhanced sSFRs relative to their matched controls, with offsets of approximately 0.20--0.30 dex. 
Despite covering different redshift ranges and applying different corrections, the sSFR offsets obtained using \sSFRhi\ and \sSFRcorr\ agree within their uncertainties (\autoref{fig:sf_diff}).
The offsets from the fitted SFMS seen in \autoref{fig:delta_ms} agree with the matched-control analysis, with slow-inflow galaxies lying $\sim$0.3 dex above the relation, corresponding to $\sim$0.8$\sigma_{\rm MS}$. 
The similarity between the slow-inflow and low-velocity blueshifted samples may indicate overlap between these two classifications. 
We return to the physical interpretation of the slow-inflow population in \autoref{sec:discuss}.

\subsubsection{Fast inflows: $v_{\rm flow} \geq 100$ \kms}
Galaxies hosting fast inflows show smaller sSFR differences than the slow-inflow population. 
Both \sSFRhi and \sSFRcorr indicate a mild enhancement of approximately 0.1 dex relative to the matched controls. 
The fitted SFMS gives the same weak enhancement: the fast-inflow population lies $\sim$0.1 dex above the relation, or $\sim$ 0.25$\sigma_{\rm MS}$. 
Thus, unlike the slow-inflow population, fast inflows are not strongly associated with elevated global star-formation activity.

This weaker enhancement may indicate that the fast-inflow sample traces a different physical regime from the slow-inflow population. 
As discussed in \citetalias{PaperI}, these high-velocity inflows may be associated with rapidly infalling gas, potentially linked to mergers or tidal interactions. 
If much of this gas is still located in the CGM rather than the inner regions of the disc, its connection to the current SFR may be weaker. 
The modest enhancement could then reflect the subset of systems where the inflowing material has already reached the galaxy centre or where dynamical disturbances have triggered star formation.

\subsection{Mass-metallicity relation}
\begin{figure*}
\centering
\includegraphics[width=\linewidth]{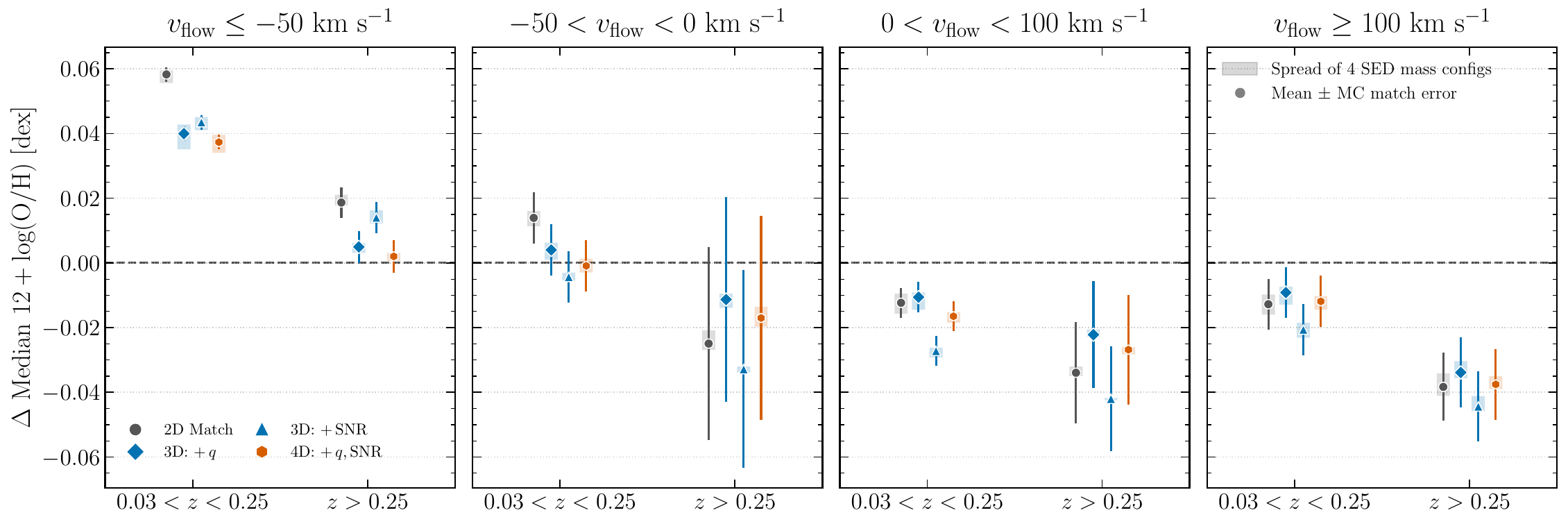}
\caption{
    Metallicity differences between galaxies hosting gas flows and matched controls, assuming metallicity depends on redshift and stellar mass.
    Plotting conventions are defined in \autoref{fig:sf_diff}.
    Metallicities are derived using the \citet{Scholte2024} strong-line calibration.
    Results are shown for galaxies at $0.03 < z < 0.25$, where the DESI fibre primarily probes the central regions, and at $z > 0.25$, where it samples a larger fraction of the disc.
    }
\label{fig:z_diff_mzr}
\end{figure*}

\begin{figure*}
\centering
\includegraphics[width=\linewidth]{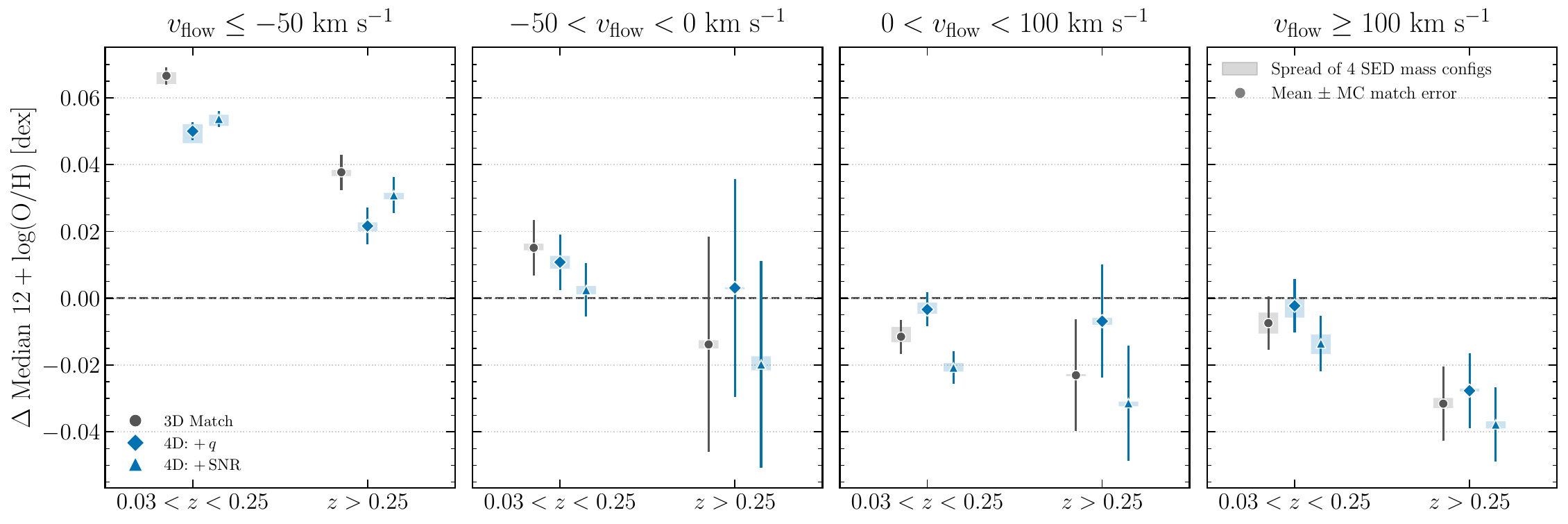}
\caption{
    Metallicity differences between galaxies with gas flows and matched control samples when control galaxies are matched in redshift, stellar mass and star-formation rate. 
    Plotting conventions are as defined in \autoref{fig:sf_diff} and \autoref{fig:z_diff_mzr}. 
    }
    \label{fig:z_diff_sfr}
\end{figure*}

We measure metallicity differences relative to matched controls and to the fitted MZR.
The resulting median metallicity differences between the gas-flow and control samples are shown in \autoref{fig:z_diff_mzr} when matching in redshift and stellar mass, and in \autoref{fig:z_diff_sfr} when additionally matching in SFR as an empirical FMR-like control.
Unlike SFRs, metallicities cannot be straightforwardly aperture corrected, so we report the matched-control comparison separately for galaxies at $0.03 < z < 0.25$ and $0.25 \leq z < 0.45$.
These redshift intervals probe different fractions of each galaxy within the DESI fibre, but they also sample different parts of the stellar-mass distribution.
In particular, the high-redshift sample is restricted to very massive galaxies, with $M_* \gtrsim 10^{11}\,\mathrm{M_\odot}$, where the MZR is relatively flat and the dependence of metallicity residuals on SFR is expected to be weak.
The lower-redshift sample extends closer to, and below, the knee of the MZR, where SFR-dependent metallicity scatter is more important.
Differences between the two redshift intervals should therefore be interpreted as reflecting both aperture coverage and the changing mass range of the selected galaxies.

\begin{figure}
\centering
\includegraphics[width=\linewidth]{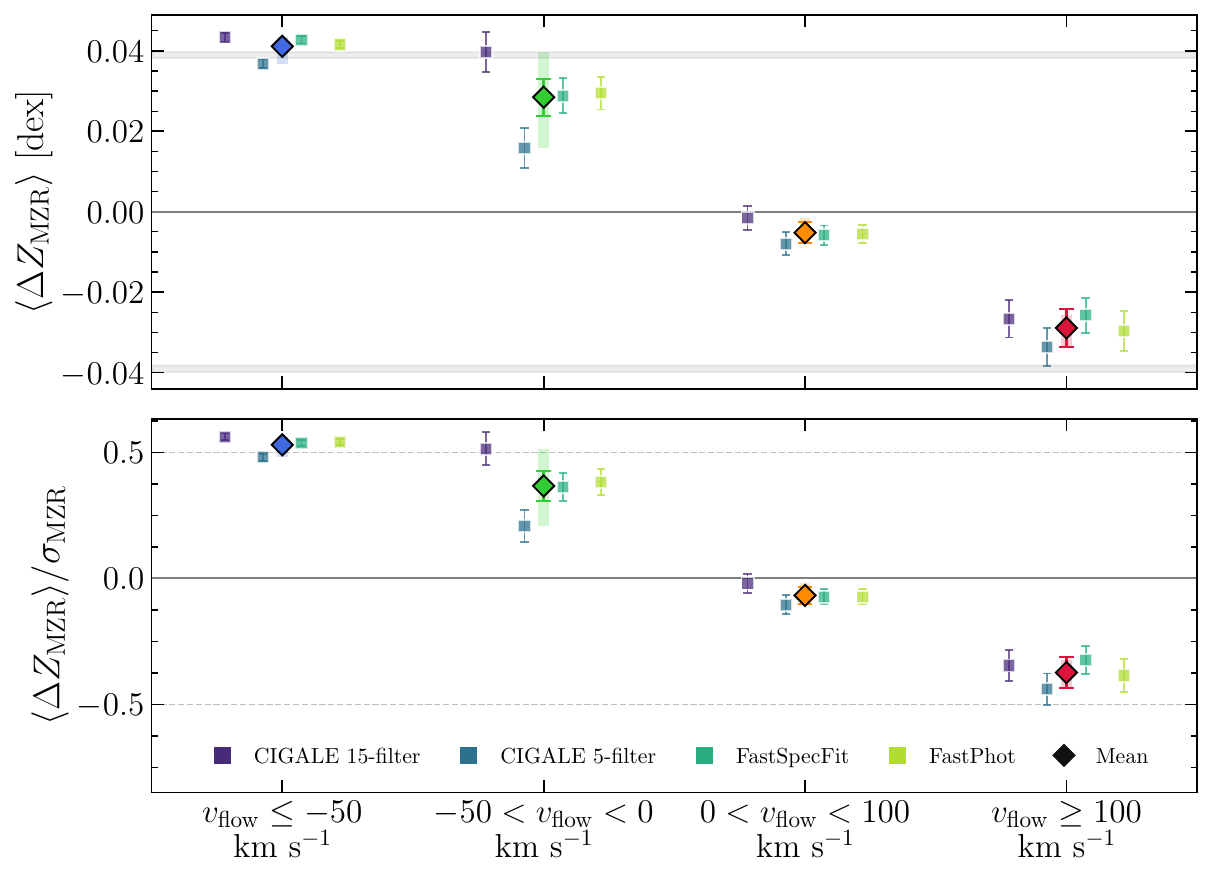}
\caption{
    Mean metallicity offset from the fitted mass–metallicity relation (MZR) for galaxies hosting gas flows at different velocities. 
    Symbols and bars are defined as in Fig.~\ref{fig:delta_ms}. 
    The upper panel presents the offset in dex, with grey bands indicating the typical ranges corresponding to \(\pm0.5\sigma_{\rm MZR}\). 
    The lower panel shows the offset normalised by the measured MZR scatter, with dashed horizontal lines at \(\pm0.5\sigma_{\rm MZR}\).
}
\label{fig:delta_mzr}
\end{figure}

In addition to the matched-control comparison, we fit the MZR for the DESI control sample and measure the residual metallicity offsets of galaxies hosting gas flows.
An example fit is shown in \autoref{fig:mzr_cg15} using \texttt{CIGALE\_5} stellar masses, with the corresponding offsets summarised in \autoref{fig:delta_mzr}.
The coloured bars show the range obtained across the different stellar-mass estimates, while the larger diamonds indicate the mean across all estimates.

We caution that the measured offsets are small compared to the systematic uncertainties associated with strong-line metallicity calibrations. 
Typical offsets between different calibrations or relative to $T_e$-based measurements can reach $\sim$0.2--0.3~dex \citep{Nakajima2022, Scholte2024}. 
We do not include these systematic uncertainties in the error budget. 
The comparison remains internally self-consistent because the gas-flow and control samples use the same calibration. 
The trends also persist across the stellar-mass estimates.

\subsubsection{Outflows: $v_{\rm flow} \leq -50$ \kms}
Galaxies with detected outflows show the clearest metallicity difference. 
In the redshift and stellar-mass matched comparison (\autoref{fig:z_diff_mzr}), the offsets are positive in both redshift bins, with $\Delta Z = +0.04$--$0.06$ dex for $0.03 < z < 0.25$ and $0.00$--$0.02$ dex for $z > 0.25$. 
The smaller offset in the higher-redshift bin may reflect the DESI fibre sampling a larger fraction of each galaxy, diluting a centrally concentrated enrichment signal. 
However, the higher-redshift bin is also restricted to the most massive galaxies, where the MZR is flatter and metallicity residuals may respond less strongly to changes in feedback.
When SFR is also included in the matching (\autoref{fig:z_diff_sfr}), the enhancement increases as expected. 
The fitted-MZR residuals in \autoref{fig:delta_mzr} support the same picture: outflow galaxies lie above the parent-sample MZR by $\sim$0.04 dex across the stellar-mass estimates, corresponding to roughly $0.5\sigma$ of the intrinsic MZR scatter. 

\subsubsection{Low-velocity blueshifted gas: $-50 < v_{\rm flow} < 0$ \kms}
Galaxies with low-velocity blueshifted absorbers show only weak evidence for metallicity differences. 
In \autoref{fig:z_diff_mzr}, the offsets in the $0.03 < z < 0.25$ bin are consistent with no difference, while those in the $z > 0.25$ bin are mildly negative, ranging from $-0.03$ to $-0.01$ dex, with large uncertainties. 
Including SFR in the matching (\autoref{fig:z_diff_sfr}) reduces the difference in the higher-redshift bin. 
The fitted-MZR residuals in \autoref{fig:delta_mzr} are more positive, $\sim$0.02--$0.04$ dex, suggesting that this population may lie slightly above the global MZR, although the matched-control comparison does not show a consistently significant enrichment. 

Small differences between the matched-control offsets and the fitted-MZR residuals are expected because the two measurements are not identical. 
The matched-control comparisons in \autoref{fig:z_diff_mzr} and \autoref{fig:z_diff_sfr} are local, non-parametric comparisons performed separately in redshift bins, whereas \autoref{fig:delta_mzr} measures residuals relative to a single fitted MZR for the parent sample. 
The latter therefore depends on the adopted functional form, mass range and the distribution of each gas-flow sample across stellar mass and redshift.

\subsubsection{Slow inflows: $0 < v_{\rm flow} < 100$ \kms}
Galaxies with slow inflows show little evidence for a strong metallicity offset. 
The direct matched comparison gives mildly negative values in both redshift bins, ranging from $-0.03$ to $-0.01$ dex for $0.03 < z < 0.25$ and from $-0.04$ to $-0.02$ dex for $z > 0.25$ (\autoref{fig:z_diff_mzr}). 
Including SFR weakens the negative offset, giving values closer to zero in both redshift bins (\autoref{fig:z_diff_sfr}). 
The MZR residuals are also consistent with no significant deviation for galaxies hosting slow inflows in \autoref{fig:delta_mzr}.

\subsubsection{Fast inflows: $v_{\rm flow} \geq 100$ \kms}
Galaxies hosting fast inflows show the clearest indication of reduced metallicity among the absorber populations. 
In the redshift and stellar-mass matched comparison (\autoref{fig:z_diff_mzr}), the offsets are negative in both redshift bins, ranging from $-0.02$ to $-0.01$ dex for $0.03 < z < 0.25$ and from $-0.04$ to $-0.03$ dex for $z > 0.25$. 
Including SFR in the matching reduces the magnitude slightly, but the offsets in the $z > 0.25$ bin remain negative at $\sim -0.04$ to $-0.03$ dex (\autoref{fig:z_diff_sfr}).

The fitted-MZR residuals in \autoref{fig:delta_mzr} are consistent with this interpretation, with fast inflow hosts lying below the parent-sample MZR by $\sim -0.03$ dex, or roughly $-0.4\sigma$ of the intrinsic scatter. 
Although modest, the agreement in sign between the matched-control and MZR-residual approaches supports the view that the fastest redshifted absorbers are associated with metal-poorer galaxies.

\subsection{Difference in 4000 \AA\, break strengths}
\begin{figure}
\centering
\includegraphics[width=\linewidth]{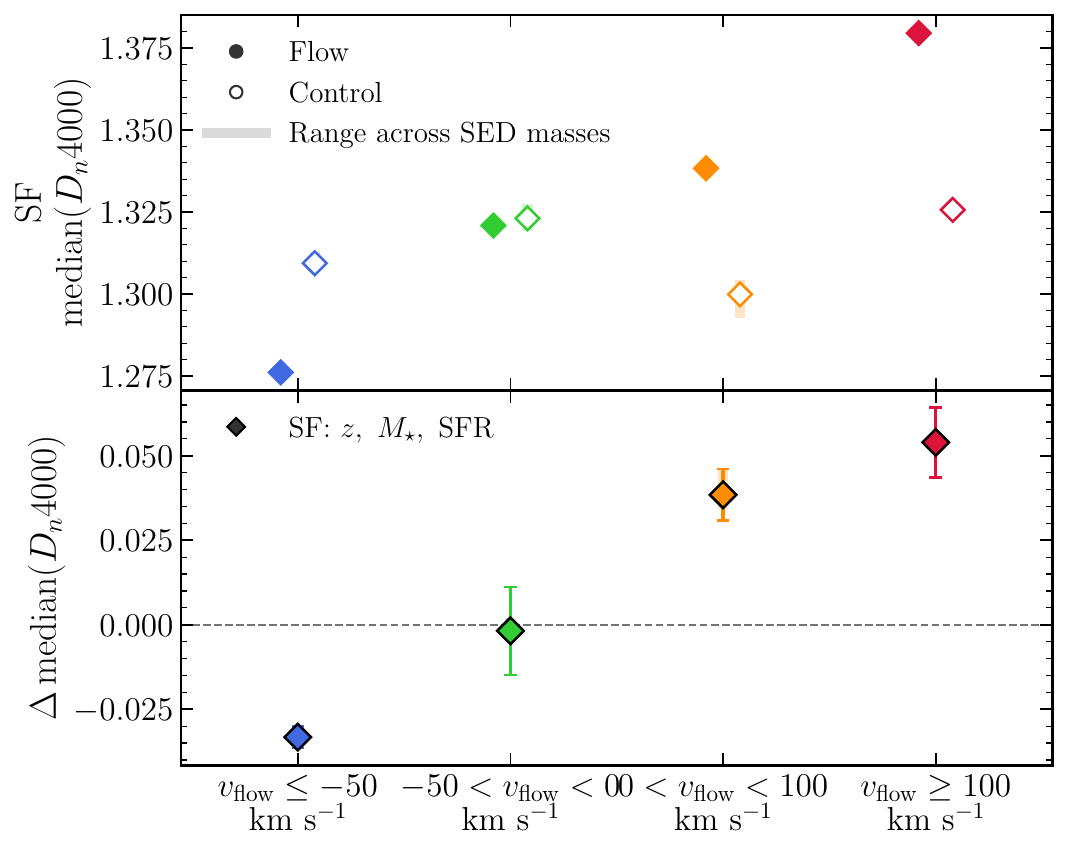}
\caption{
    Comparison of median $D_n4000$ measurements for star-forming galaxies hosting gas flows and matched star-forming controls.
    The top panel shows the absolute median $D_n4000$ values for the flow and control samples, while the bottom panel shows the offset between them, defined as $D_n4000_{\rm flow}-D_n4000_{\rm control}$.
    Markers show the mean across the four stellar-mass estimates, while vertical bars span their full range; error bars in the offset panel show the typical uncertainty for an individual estimate.
    Negative (positive) offsets indicate weaker (stronger) 4000\,\AA\ breaks relative to the matched controls. 
    }
\label{fig:dn4000}
\end{figure}

We compare the median $D_n4000$ of star-forming galaxies hosting gas flows with that of matched star-forming control samples in \autoref{fig:dn4000}. 
The top panel shows the absolute median $D_n4000$ values for the flow and control samples, while the bottom panel shows the corresponding offsets, defined as $D_n4000_{\rm flow}-D_n4000_{\rm control}$, such that negative values indicate weaker breaks in the gas-flow sample. 
The controls are matched in redshift, stellar mass and SFR, testing whether gas-flow hosts differ in their recent star-formation histories after controlling for their current SFR. 
The markers show the mean value obtained across the four stellar-mass estimates, while the vertical bars show the variation between them.

The absolute values in the top panel show that both the flow and control samples have median $D_n4000 < 1.4$, consistent with relatively young, actively star-forming stellar populations within the fibre. 
The $D_n4000$ offsets are small but vary systematically with gas-flow velocity after matching in redshift, stellar mass and SFR. 
Outflow hosts retain a negative offset, indicating weaker 4000\,\AA\ breaks than their matched controls even at fixed SFR. 
The difference for the low-velocity blueshifted population is close to zero. 
The inflow populations instead show positive offsets, with slow and fast inflows having slightly stronger breaks than their matched controls. 
This suggests that their current star formation may be occurring on top of a somewhat older underlying stellar population.
The absolute medians nevertheless remain below 1.4, so the positive offsets represent small relative differences within an entirely star-forming population rather than evidence that the inflow hosts were previously quiescent. 
As $D_n4000$ is sensitive to both age and metallicity and is measured through a single aperture, these trends should be interpreted as relative differences in the central stellar populations rather than as pure age differences or constraints on the galaxies as a whole.

\section{Discussion}
\label{sec:discuss}

\subsection{Summary of the observational constraints}
The SFR, gas-phase metallicity and $D_n4000$ measurements probe different components and timescales of the galaxies hosting the detected gas flows. 
The H$\alpha$-based SFR traces star formation on 10--20 Myr timescales, gas-phase metallicity measures the present chemical state of the ionised ISM and $D_n4000$ is sensitive to the longer-term stellar population. 
They likely do not respond on similar timescales to the same gas-flow event.

Outflow hosts show enhanced sSFRs, elevated metallicities and slightly weaker central 4000 \AA\ breaks than their matched controls. 
Slow-inflow hosts also have elevated sSFRs and no significant metallicity difference, while their slightly stronger $D_n4000$ indicates a modest difference from the recent stellar populations of the controls. 
However, the absolute $D_n4000$ values remain characteristic of star-forming galaxies.
Fast-inflow hosts show only a modest sSFR enhancement but evidence for lower gas-phase metallicity and slightly stronger $D_n4000$ than the control sample. 
The low-velocity blueshifted population has enhanced sSFR, with a slight metallicity enhancement and no $D_n4000$ difference.

\subsection{Physical interpretation of the gas-flow classes}
\subsubsection{Outflows during elevated star formation}
In the local Universe, galactic outflows are commonly associated with regions where the star-formation surface densities exceed $\Sigma_{\rm SFR} > 0.1\ \rm M_\odot\ yr^{-1}\ kpc^{-2}$ \citep{Heckman2002}. 
A substantial fraction of the parent sample also exceeds this threshold, however, so elevated star-formation surface density alone is not sufficient to guarantee a detectable down-the-barrel outflow. 
Controlling for axial ratio leaves the sSFR enhancement largely unchanged and hence, viewing geometry alone does not explain the difference. 
The outflow sample is therefore likely weighted towards stronger winds that are more readily detected or produced during periods of elevated star formation.

The lower central $D_n4000$ of outflow hosts, including the residual difference after matching in SFR, is consistent with a modestly greater contribution from young stars to the fibre continuum than in the controls. 
This may reflect recent central star formation extending over a longer interval than that traced by H$\alpha$, but the small offset and the dependence of $D_n4000$ on stellar metallicity prevent a unique constraint on the age or duration of the episode. 
If young stars are the dominant contributor to the offset, they may trace sustained recent central star formation that supplies the supernova and stellar-wind feedback required to drive a stronger outflow. 

Galactic winds are expected to reduce gas-phase metallicity over time by ejecting metal-enriched material from the interstellar medium and thereby lowering the effective yield of the galaxy \citep{Chisholm2018}.
This effect is strongest when winds are preferentially enriched by recently synthesised metals or when the expelled gas is not rapidly recycled.
Simulations similarly find that mass-dependent metal loss and recycling play an important role in shaping the MZR, with lower-mass galaxies generally retaining a smaller fraction of the metals they produce \citep{Finlator2008, Dave2011, Peeples2011, vanLoon2021}.
The observed metallicity enhancement therefore suggests that outflow hosts are caught during a chemically enriched phase of the central ISM, but its modest amplitude, reaching only $\sim$0.5$\sigma_{\rm MZR}$ above the MZR, may indicate that some enriched material has already been expelled into the CGM.
Subsequent removal of additional enriched gas, together with mixing and recycling, could then return the galaxy toward the mean MZR.

\subsubsection{Slow and fast inflows}
The elevated SFRs and absence of a metallicity deficit in slow-inflow hosts do not resemble a simple instantaneous response to the first arrival of a large mass of metal-poor gas. 
The result could arise if the inflow is enriched, if the accreted mass is small relative to the existing reservoir or if an earlier dilution signal has been reduced by mixing and ongoing enrichment. 
As \ion{Na}{i} D predominantly traces neutral gas, a detected inflow may not represent gas that has already reached the molecular phase or is immediately available for star formation \citep{Langan2023}. 
Slow inflows may preferentially occur in gas-rich systems whose higher gas fractions are themselves associated with greater star-formation activity \citep{Saintonge2022}. 
The slightly stronger central $D_n4000$ measured in slow-inflow hosts adds a further nuance: their current star formation appears to be superimposed on a somewhat older stellar population than in the controls. 
However, the absolute median remains below 1.4, so this result does not imply that these galaxies were previously quiescent or that globally quenched, gas-poor systems have restarted star formation. 

An alternative interpretation is that the enhanced star formation precedes, rather than follows, the observed inflow.
In a galactic-fountain scenario, star formation and feedback can lift metal-enriched material out of the disc, after which some fraction cools, stalls or rains back onto the galaxy as low-velocity neutral gas \citep{Fraternali2008, Marinacci2010, Fraternali2017}.
Such recycled accretion would naturally produce inflow hosts with elevated SFRs but little or no metallicity dilution. 
In this case, the observed slow inflows would trace feedback-driven recycling.

A secular origin may also contribute to the slow-inflow population. 
Bars can drive radial transport and enhance central star formation by approximately the magnitude measured here \citep{Ellison2011, Chown2019, Fraser-McKelvie2020, Geron2021}. 
They can also funnel enriched gas towards galaxy centres \citep{Friedli1994}, although the absence of a metallicity enhancement in the slow-inflow sample does not provide direct support for this part of the bar interpretation. 
Other studies find unchanged or suppressed star formation in massive barred galaxies \citep{Vera2016, Maeda2023, Liu2026}. 
Bars are therefore one plausible contributor rather than a complete explanation, and testing this possibility requires direct morphological measurements.

Fast inflow hosts show a different combination of properties: only a small current sSFR enhancement, lower gas-phase metallicities and a stronger $D_n4000$.
This pattern is consistent with metal-poor material having reached or begun mixing with the disc before producing a strong star-formation response.
The metallicity decrease is stronger at higher redshift, where the DESI fibre subtends a larger fraction of the galaxy, suggesting that the dilution may extend beyond the central regions sampled at low redshift \citep{Cameron2021}.
The slightly stronger central $D_n4000$, together with its absolute value in the star-forming regime, is consistent with a star-formation response that has not become strong enough to dominate the recent fibre continuum; it does not establish an old global stellar population.
As discussed in \citetalias{PaperI}, this class may include gas associated with mergers, tidal structures or rapidly infalling CGM material.
The detected neutral gas may consequently be less directly coupled to the current star-forming disc.

\subsection{Illustrative enrichment and dynamical timescales}
The interpretations above all depend on the relative timing of gas accretion, mixing, enrichment, star formation and gas expulsion.
The combination of enhanced sSFR, no significant metallicity deficit and stronger $D_n4000$ in slow-inflow hosts may therefore reflect a mismatch between the timescales traced by each observable, rather than a single instantaneous response to inflow.
In particular, the absence of a metallicity offset does not rule out an earlier dilution event.
For a gas reservoir with metallicity $Z$, depletion time $t_{\rm dep}=M_{\rm gas}/{\rm SFR}$ and effective yield $y_{\rm eff}$, an illustrative recovery time from a deficit of magnitude $\Delta\log Z$ is
\begin{equation}
    t_{\rm enrich} \sim
    \frac{\left(10^{\Delta\log Z}-1\right)Z}{y_{\rm eff}}\,t_{\rm dep}.
\end{equation}
For example, gas with $0.3Z$ contributing 10\% of the pre-existing gas mass would initially produce approximately 0.03 dex of dilution.
Representative oxygen yields \citep{Kudritzki2015} and an assumed molecular-gas depletion time of 1 Gyr for local massive galaxies \citep{Saintonge2011} give a $t_{\rm enrich} \approx$ 0.1--0.3 Gyr.
A small metallicity deficit could thus return to the control value on a sub-Gyr timescale, while the galaxy still shows enhanced star formation or an observable neutral inflow.
This estimate is not a measurement of the time elapsed since inflow because it depends strongly on the inflow metallicity and mass, mixing, gas exchange, metal retention and star-formation history.

The characteristic inflow velocities provide a complementary dynamical scale.
For a given inflow speed, $v_{\rm inflow}$, the time required to move across a distance $R$ is
\begin{equation}
    t_{\rm cross} \sim 0.98~{\rm Gyr}
    \left(\frac{R}{1~{\rm kpc}}\right)
    \left(\frac{v_{\rm inflow}}{1~{\rm km~s^{-1}}}\right)^{-1}.
\end{equation}
Gas moving over $R\sim 5$--10 kpc at the typical slow-inflow speed of 10--20 km s$^{-1}$ would therefore require $t_{\rm cross}\sim 0.25$--1 Gyr, much longer than the $\sim$10 Myr timescale traced by H$\alpha$.
The estimate is highly geometry dependent because the measured velocity is a line-of-sight component and the location of the absorbing gas is uncertain.
Even with this uncertainty, slow \ion{Na}{i}~D inflows can trace long-lived transport during which shorter episodes of star formation, enrichment and mixing occur.

\subsection{A possible regulatory cycle}
The positions of the gas-flow samples within the SFMS scatter motivate, but do not demonstrate, a regulatory sequence. 
Slow-inflow hosts lie $\sim$0.8$\sigma_{\rm MS}$ above the SFMS and outflow hosts lie approximately $1\sigma_{\rm MS}$ above it. 
These offsets describe the locations of the selected populations; they do not measure the fraction of the total SFMS scatter caused by gas flows. 
That contribution depends on the true incidence and duration of gas-flow phases, neither of which is constrained by the observed offsets.

A possible sequence, consistent with gas-regulator or ``bathtub'' models, is that inflows increase the available gas reservoir and move galaxies above the SFMS, after which gas consumption and feedback-driven outflows remove gas and help regulate the enhanced activity \citep[e.g.][]{Bouche2010, Lilly2013, Forbes2014, Tacchella2016}.
The chemical measurements fit a related picture: fast inflows produce dilution, mixing and enrichment then erase the deficit, and outflows become visible while the central ISM is still metal rich \citep{Finlator2008, Dave2012}.
At low redshift, these measurements describe primarily the central gas-phase metallicity because the DESI fibre samples only the inner region of each galaxy.
If metal-poor inflowing gas first mixes with the ISM at larger galactocentric radii, dilution could occur outside the fibre before reaching the central region.
An earlier or spatially extended drop below the global MZR could consequently escape detection.
The $D_n4000$ results further allow the possibility that inflows are associated with recently enhanced star formation superimposed on a somewhat older stellar population than in the controls. The small offsets and single-aperture measurements, however, do not uniquely establish the preceding star-formation history.

There are roughly 200 systems in which outflows and slow inflows are detected simultaneously \citepalias{PaperI}, demonstrating that inflows and outflows can occur concurrently.
The observable gas motions and the star-formation response is not necessarily tightly synchronised and individual galaxies do not necessarily pass through the stages in a fixed order.
The interpretation becomes even less unique if some slow inflows trace galactic fountains, because the observed redshifted gas may then be recycled material launched by earlier star formation rather than newly accreted gas driving the current activity.
In that case, the causal ordering between inflow, star formation and feedback is ambiguous.

\subsection{Future work}
\begin{figure*}
\centering
\includegraphics[width=\linewidth]{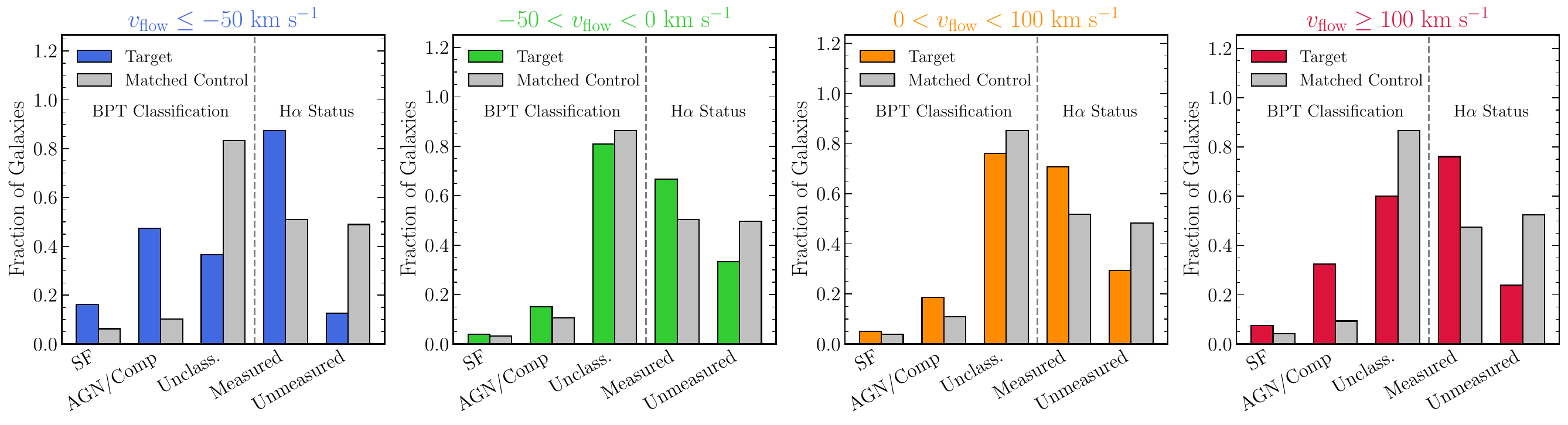}
\caption{
    Fraction of galaxies belonging to different [\ion{N}{ii}] BPT \citep{Kewley2001, Kauffmann2003} and H$\alpha$ detection classes for the gas-flow populations and their stellar mass and redshift matched control samples. 
    From left to right, the coloured histograms show the fractions for outflows, low-velocity blueshifted systems, slow inflows and fast inflows, while the grey histograms represent the corresponding control samples. 
    The dashed line separates the BPT classifications from the H$\alpha$ detection statistics. 
    }
\label{fig:frac}
\end{figure*}

The primary scaling-relation analysis is restricted to star-forming galaxies with reliable SFRs and BPT classifications. 
The restriction gives a uniform comparison sample but excludes much of the gas-flow population. 
\autoref{fig:frac} shows that the majority of gas-flow galaxies with a BPT classification host AGN and that a non-zero fraction lack H$\alpha$ emission with SNR $> 3$.

Approximately 15\% of outflow hosts and $\sim$30\% of inflow hosts do not have measurable H$\alpha$, although both fractions remain below those of the corresponding controls. 
This again emphasises that gas flows and star formation are not perfectly synchronised. 
Inflows may precede detectable star formation while neutral gas settles and condenses, whereas outflows may remain observable after the most recent H$\alpha$-bright episode has faded.

The dominance of AGN among the classified gas-flow population sets a broader limit on the interpretation: the star-formation-centred picture developed for the selected sample cannot automatically be generalised to the majority of inflow and outflow hosts. 
AGN may drive some flows directly or alter the coupling between gas transport and star formation \citep{Davies2024}. 
Jointly analysing AGN-driven and star-formation-driven flows is therefore a central requirement for interpreting the full baryon cycle.

Future work will also move beyond single-aperture, instantaneous diagnostics.
Full spectral fitting with tools such as \texttt{pPXF} \citep{PPXF2004, PPXF2017} or \texttt{Bagpipes} \citep{Carnall2018} would distinguish sustained, declining and episodic star-formation histories more directly than $D_n4000$ alone.
Measurements of mass inflow and outflow rates, together with the total gas reservoir, are needed to establish whether the detected inflows can sustain the observed star formation and whether the subsequent outflows can significantly deplete that reservoir.
Spatially resolved spectroscopy would further test whether metallicity dilution occurs outside the DESI fibre before reaching the central regions and would help separate radial accretion, galactic fountains and secular transport.

\section{Conclusions}
\label{sec:conclusion}
Using more than 6,000 star-forming galaxies with down-the-barrel \ion{Na}{i}~D absorption from DESI DR2, we investigated where systems with detected neutral-gas inflows and outflows lie relative to the SFMS and MZR.
We compared galaxies with inflows and outflows to matched control samples and measured their offsets from scaling relations fitted to the DESI control sample. 
Our main conclusions are as follows.

\begin{itemize}
    \item Galaxies hosting outflows ($v_{\rm flow} \leq -50$ \kms) show enhanced star-formation activity relative to matched controls, with sSFR offsets of $\sim$0.25--0.40 dex. 
    They are also metal enhanced by $\sim$0.04--0.06 dex in the lower-redshift sample. 
    Their slightly weaker central 4000\,\AA\ breaks, after matching in SFR, are consistent with a greater contribution from young stars than in the controls.

    \item Galaxies with slow inflows ($0 < v_{\rm flow} < 100$ \kms) are also elevated above the SFMS, with sSFR enhancements of $\sim$0.20--0.30 dex. 
    They show no significant metallicity difference relative to matched controls. 
    After matching in SFR, they have slightly stronger central 4000\,\AA\ breaks than their controls, although the absolute values remain characteristic of star-forming populations. 
    This modest difference is consistent with current star formation being superimposed on a somewhat older stellar population than in the controls, but may also reflect differences in stellar metallicity.

    \item Fast inflows ($v_{\rm flow} \geq 100$ \kms) show much smaller sSFR offsets and marginal evidence for reduced metallicity. 
    Together with their slightly stronger central 4000\,\AA\ breaks in star-forming galaxies, this may indicate recently arriving gas that is weakly coupled to current star formation, potentially associated with mergers or tidal structures.

    \item The slow-inflow population lies $\sim$0.8$\sigma_{\rm MS}$ above the SFMS, while galaxies with outflows lie approximately $1\sigma_{\rm MS}$ above it. 
    These offsets motivate a possible regulatory cycle in which inflows increase the star-forming reservoir and move galaxies above the relation, before outflows remove gas and help reduce the enhanced activity. 
    However, inflow hosts may also simply trace gas-rich galaxies in which the conversion of neutral gas into molecular gas and then stars occurs over extended timescales, so the observed inflow is not necessarily the immediate cause of the elevated star formation.

    \item Our results highlight the importance of direct gas-flow measurements for interpreting scatter in galaxy scaling relations.
    Large surveys of down-the-barrel \ion{Na}{i}~D absorbers provide a powerful way to identify galaxies undergoing gas transport, while integral field spectroscopy will be required to connect these flows to the spatially resolved baryon cycle of gas accretion, star formation, metal enrichment and feedback.
\end{itemize}

\begin{acknowledgements}
 This material is based upon work supported by the U.S. Department of Energy (DOE), Office of Science, Office of High-Energy Physics, under Contract No. DE–AC02–05CH11231 and by the National Energy Research Scientific Computing Center, a DOE Office of Science User Facility under the same contract. Additional support for DESI was provided by the U.S. National Science Foundation (NSF), Division of Astronomical Sciences under Contract No. AST-0950945 to the NSF’s National Optical-Infrared Astronomy Research Laboratory; the Science and Technology Facilities Council of the United Kingdom; the Gordon and Betty Moore Foundation; the Heising-Simons Foundation; the French Alternative Energies and Atomic Energy Commission (CEA); the Secretariat of Science, Humanities, Technology and Innovation (SECIHTI) of Mexico; the Ministry of Science, Innovation and Universities of Spain (MICIU/AEI/10.13039/501100011033) and by the DESI Member Institutions: \url{https://www.desi.lbl.gov/collaborating-institutions}.
\newline
The DESI Legacy Imaging Surveys consist of three individual and complementary projects: the Dark Energy Camera Legacy Survey (DECaLS), the Beijing-Arizona Sky Survey (BASS) and the Mayall z-band Legacy Survey (MzLS). DECaLS, BASS and MzLS together include data obtained, respectively, at the Blanco telescope, Cerro Tololo Inter-American Observatory, NSF’s NOIRLab; the Bok telescope, Steward Observatory, University of Arizona; and the Mayall telescope, Kitt Peak National Observatory, NOIRLab. NOIRLab is operated by the Association of Universities for Research in Astronomy (AURA) under a cooperative agreement with the National Science Foundation. Pipeline processing and analyses of the data were supported by NOIRLab and the Lawrence Berkeley National Laboratory. Legacy Surveys also uses data products from the Near-Earth Object Wide-field Infrared Survey Explorer (NEOWISE), a project of the Jet Propulsion Laboratory/California Institute of Technology, funded by the National Aeronautics and Space Administration. Legacy Surveys was supported by: the Director, Office of Science, Office of High Energy Physics of the U.S. Department of Energy; the National Energy Research Scientific Computing Center, a DOE Office of Science User Facility; the U.S. National Science Foundation, Division of Astronomical Sciences; the National Astronomical Observatories of China, the Chinese Academy of Sciences and the Chinese National Natural Science Foundation. LBNL is managed by the Regents of the University of California under contract to the U.S. Department of Energy. The complete acknowledgements can be found at \url{https://www.legacysurvey.org/}.
\newline
Any opinions, findings and conclusions or recommendations expressed in this material are those of the author(s) and do not necessarily reflect the views of the U. S. National Science Foundation, the U. S. Department of Energy or any of the listed funding agencies.

The authors are honored to be permitted to conduct scientific research on I'oligam Du'ag (Kitt Peak), a mountain with particular significance to the Tohono O’odham Nation. 

\newline
This work was supported by the French National Research Agency (ANR) under contract ANR-22-CE31-0026.

\newline
We thank Rebecca Davies and Sara Ellison for helpful discussions that clarified the interpretation of our results, and Abigail Bault and Zhiwei Shao for constructive comments that improved the manuscript.
\end{acknowledgements}

\section*{Data Availability}
The data used in this analysis will be made public after Data Release 2 (details in \url{https://data.desi.lbl.gov/doc/releases/}). 
The data corresponding to the figures in this paper will be made publicly available in a Zenodo repository (\url{https://zenodo.org/records/21992757}).

\bibliographystyle{aa} % style aa.bst
\bibliography{bib,DESI} % your references Yourfile.bib

\appendix
\section{SED fitting codes}
\label{app:SED}
We use four different stellar-mass estimates to assess the systematic uncertainty associated with the choice of SED-fitting code and input data. 
Our analysis relies primarily on relative comparisons between gas-flow galaxies and control samples, rather than on the absolute accuracy of any individual mass estimate. 
For a given redshift and set of photometric measurements, each code should provide internally self-consistent masses; repeating the analysis with all four estimates therefore tests whether our conclusions depend on the adopted SED-fitting methodology.

{All four estimates assume a \citet{Chabrier2003} IMF. 
\texttt{FastSpecFit} uses solar-metallicity FSPS templates and a five-bin non-parametric star-formation history to fit the DESI spectroscopy and broadband photometry, while \texttt{FastPhot} applies the same models to the photometry alone. 
The CIGALE$_{15}$ and CIGALE$_5$ estimates \citep{ZouCIGALE2024} both use \citet{Bruzual2003} stellar populations, five stellar metallicities and a delayed star-formation history. 
CIGALE\_15 fits five broadband photometric bands together with ten synthetic bands derived from the DESI spectrum, while CIGALE\_5 uses only the five broadband bands.}

\section{Example matching and scaling-relation figures}
\label{app:example_figures}
\begin{figure*}
\centering
\includegraphics[width=\linewidth]{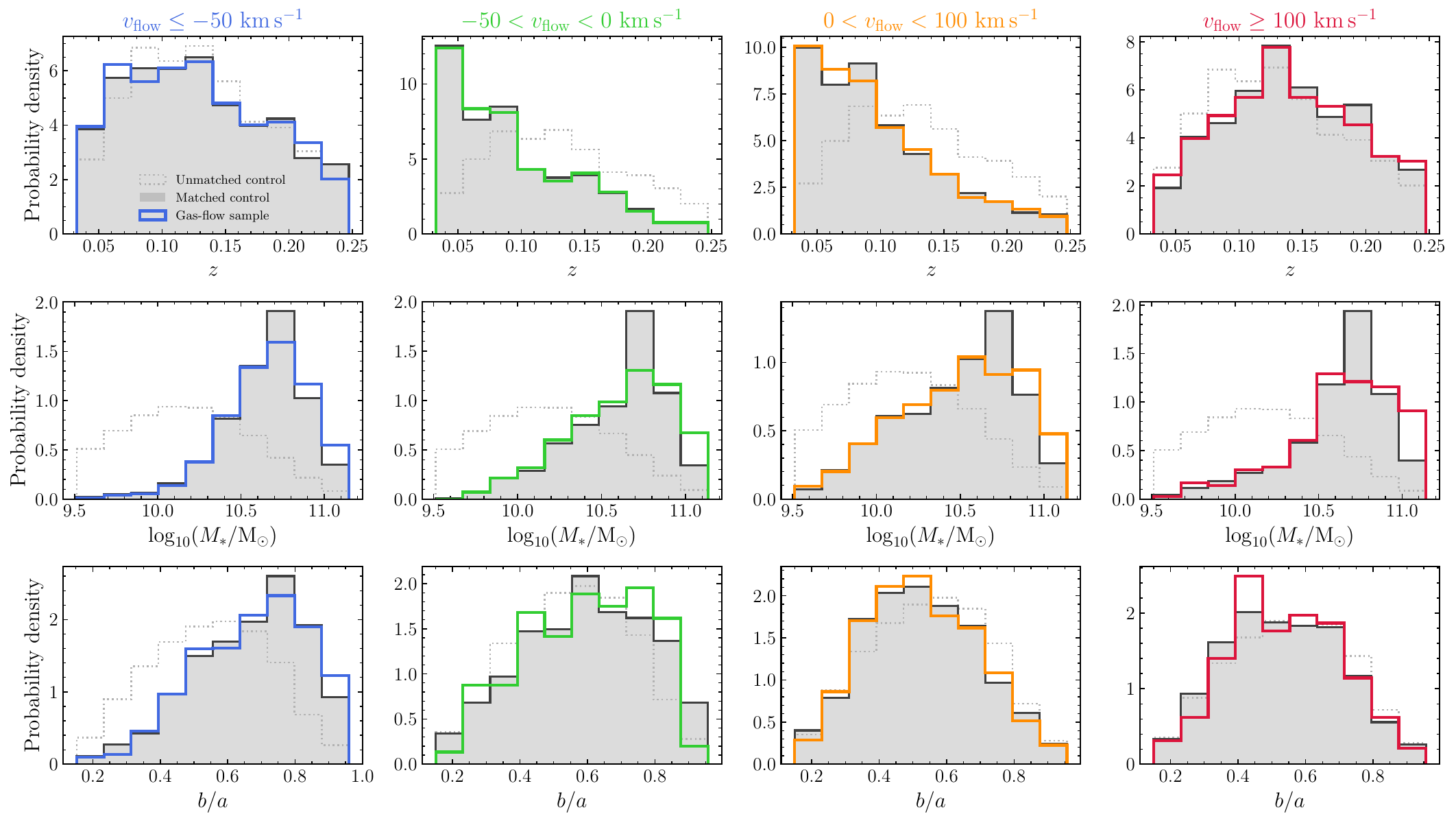}
\caption{
    Three-parameter matching diagnostic corresponding to \autoref{fig:hist_show}. 
    Normalised redshift (top), stellar-mass (middle) and axial-ratio (bottom) distributions for the four gas-flow categories. 
    Coloured lines show the gas-flow samples, solid black lines and grey shading show the matched control samples and dotted grey lines show the parent sample. 
    Controls were matched jointly in redshift, stellar mass and axial ratio using 10 bins per dimension. 
    }
\label{fig:match}
\end{figure*}

\autoref{fig:match} illustrates the matching quality for one representative configuration. 
Although the unmatched parent controls can differ substantially from the gas-flow samples, particularly in stellar mass, the matched controls closely reproduce the redshift, stellar-mass and axial-ratio distributions of each flow class. 
This agreement reduces the likelihood that the measured sSFR differences are driven by residual differences in these matching variables rather than by the presence of detected gas flows.

\begin{figure*}
\centering
\includegraphics[width=\linewidth]{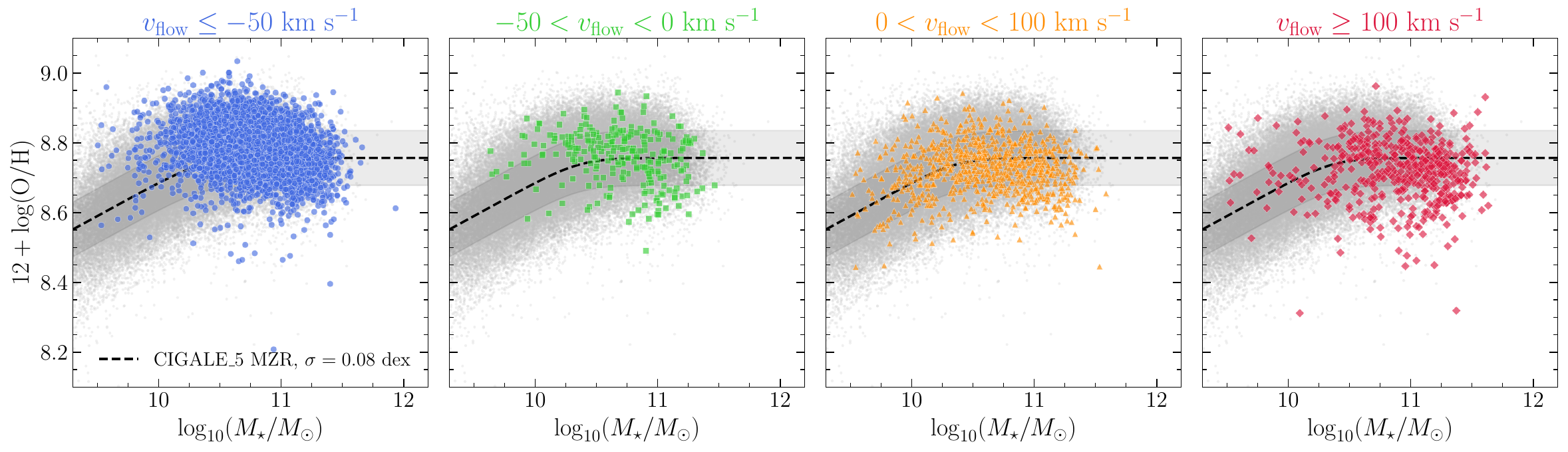}
\caption{
    The mass-metallicity relation for galaxies with different gas-flow velocities. 
    In this example, metallicities are calculated using the \citet{Scholte2024} calibration and stellar masses are taken from the \texttt{CIGALE\_5} estimates. 
    The dashed line shows the best-fitting MZR derived from the parent sample, with the grey shaded region showing the intrinsic $1\sigma$ scatter. 
    Galaxies with different gas-flow velocities are shown as coloured points in each panel. 
}
\label{fig:mzr_cg15}
\end{figure*}

\autoref{fig:mzr_cg15} shows a representative MZR fit and the locations of the four gas-flow classes relative to the parent sample. 
All four classes occupy the same broad relation as the parent population, and the class-dependent shifts are small compared with the intrinsic scatter of $0.08$ dex. 
The median residuals and matched-control offsets reported in \autoref{fig:delta_mzr} quantify these differences more directly than the visual separation of the individual points.

\end{document}